\documentclass[journal=jacsat,manuscript=article]{achemso}
\setkeys{acs}{keywords=true, articletitle=true, chaptertitle=true, doi=true}
\usepackage[version=3]{mhchem} % Formula subscripts using \ce{}

\usepackage{amssymb}
\usepackage{amsmath}
\usepackage{caption}
\usepackage{subcaption}
\usepackage{hyperref}
\usepackage{xcolor}
\usepackage{booktabs}
\usepackage{lscape}
\usepackage{natbib}

\author{Zubin Trivedi}
\email{zubin.trivedi@imsb.uni-stuttgart.de}
\affiliation[imsb-unistuttgart]{Institute for Modelling and Simulation of Biomechanical Systems, University of Stuttgart, Pfaffenwaldring 5a, 70569 Stuttgart}
\altaffiliation{These authors contributed equally to this work}
\author{Jacek K.~Wychowaniec}
\altaffiliation{These authors contributed equally to this work}
\author{Dominic Gehweiler}
\author{Christoph M.~Sprecher}
\author{Andreas Boger}
\affiliation[hs-ansbach]{Ansbach University of Applied Sciences, Residenzstraße 8, 91522 Ansbach, Germany}
\author{Boyko Gueorguiev}
\author{Matteo D'Este}
\affiliation[ari-davos]{AO Research Institute Davos, Clavadelerstrasse 8, 7270 Davos, Switzerland}
\author{Tim Ricken}
\affiliation[isd-unistuttgart]{Institute of Structural Mechanics and Dynamics in Aerospace Engineering, University of Stuttgart, Pfaffenwaldring 27, 70569 Stuttgart}
\author{Oliver R\"ohrle}
\affiliation[imsb-unistuttgart]{Institute for Modelling and Simulation of Biomechanical Systems, University of Stuttgart, Pfaffenwaldring 5a, 70569 Stuttgart}
\alsoaffiliation[simtech]{Stuttgart Center for Simulation Science (SC SimTech), Pfaffenwaldring 5a, 70569 Stuttgart}

\title{Rheological Analysis and Evaluation of Measurement Techniques for the Curing Polymethylmethacrylate Bone Cement in Vertebroplasty}

\begin{document}

\begin{tocentry}
\includegraphics[width=\linewidth]{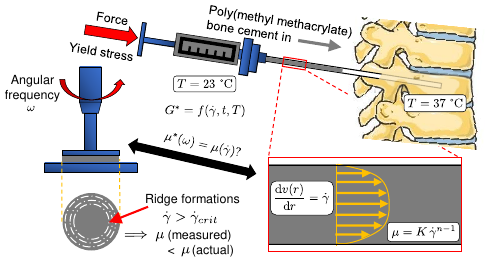}
% Some journals require a graphical entry for the Table of Contents.
% This should be laid out ``print ready'' so that the sizing of the

% Inside the \texttt{tocentry} environment, the font used is Helvetica
% 8\,pt, as required by \emph{Journal of the American Chemical

% The surrounding frame is 9\,cm by 3.5\,cm, which is the maximum
% permitted for  \emph{Journal of the American Chemical Society}
% graphical table of content entries. The box will not resize if the
% content is too big: instead it will overflow the edge of the box.

% This box and the associated title will always be printed on a
% separate page at the end of the document.

\end{tocentry}

\begin{abstract}
Vertebroplasty is a minimally invasive surgical procedure used to treat vertebral fractures, which conventionally involves injecting polymethylmethacrylate (PMMA) bone cement into the fractured vertebra. A common risk associated with vertebroplasty is cement leaking out of the vertebra during the injection, which may occur due to a lack of understanding of the bone cement's complex flow behavior. Therefore, experiments to quantify the cement's flow properties are necessary for understanding and proper handling of the bone cement. In this study, we aimed to characterize the behavior of the PMMA bone cement in its curing stages to obtain parameters that govern the flow behavior during the injection. We used rotational and oscillatory rheometry for our measurements, as well as a custom-made injector setup that replicated a typical vertebroplasty setting. Our results showed that the complex viscoelastic behavior of the bone cement is significantly affected by deformations and temperature. We found that the results from rotational tests, often used for characterizing the bone cement, are susceptible to measurement artifacts caused by wall slip and ``ridge"-like formations in the test sample. We also found the Cox-Merz rule to be conditionally valid, which affects the use of oscillatory tests to obtain shear-thinning characteristics for the bone cement. Our findings identify important differences in the measured flow behavior of PMMA bone cement when assessed by different rheological methods, understanding which is crucial for its risk-free usage in downstream medical applications.
\end{abstract}

\section{Introduction}
\label{sec:intro}
Vertebral fractures are a major source of suffering and disability, carrying risk of back pain, immobility, and potentially severe post-complications that impact the quality of life \cite{barton_rates_2019, oei_osteoporotic_2018}. Up to 28\% of patients with osteoporotic fracture-related issues resulted in death by 2008 \cite{teng_mortality_2008}. Unlike many other musculoskeletal complications where bone grafting or alternative methods can be used \cite{verdugo-avello_current_2021}, the most commonly used procedure to treat vertebral fractures is vertebroplasty \cite{galibert_percutaneous_1990, lou_percutaneous_2019, wang_balloon_2018}. In this procedure, a rapidly curing bone cement is injected into the porous cancellous part of the vertebra to serve as mechanical support \cite{bohner_theoretical_2003, robo_long-term_2021, jensen_percutaneous_1997} and allow early mobilization, especially in the elderly, where lack of mobility often causes a cascade of medical issues and often death \cite{billot_preserving_2020}. A successful vertebroplasty can provide instant pain relief \cite{mcgraw_prospective_2002} to the patient. On the other hand, improper injection could lead to cement leaking outside the vertebra, which could cause serious complications like pulmonary embolism and paralysis \cite{bernhard_asymptomatic_2003, ratliff_root_2001}. Bone cement is a material with complex rheological behavior dependent on many factors, which makes it hard to predict its behavior during injection, especially for practitioners who rely on intuition, skill, and sensory feedback. Therefore, characterization of the curing bone cement using various measurement techniques is helpful in understanding its behavior during injection in various conditions. Mathematical constitutive models derived from the characterization data can quantify the various dependencies, which can be used in analytical or computational models to reliably predict the behavior of the bone cement during injection \cite{ahmadian_digital_2022, widmer_mixed_2011, baroud_finite_2004, lepoutre_bone_2019, lian_biomechanical_2008, trivedi_continuum_2023}. 

The bone cement typically consists of ceramic components and fast-curing, minimally-toxic polymers that rapidly polymerize into, e.g.~poly(methyl methacrylate) (PMMA) \cite{bohner_theoretical_2003, dai_novel_2021, demir-oguz_injectable_2023, zhu_bioactive_2020}. The rapid polymerization (curing) of bone cement is designed to provide the required structural support for the damaged vertebra, whereas ceramic components like zirconium dioxide function as radio-opacifiers that allow for easy visualization during the vertebroplasty procedure \cite{aghyarian_vitro_2017}. Hydroxyapatite, a calcium phosphate mineral naturally found in bone, is also sometimes added to improve biocompatibility and promote bone growth around the cement. The curing typically relies on free radical polymerization, i.e.~the initiators react to form free radicals, which cause the methyl methacrylate (MMA) monomer molecules to form PMMA polymer chains \cite{ali_review_2015, hatada_stereoregular_1988}. The PMMA polymerization is an exothermic process \cite{ali_review_2015}, that is known to increase local temperatures and contribute to some surrounding tissue damage. At the same time, increased local temperature increases the polymerization rate of PMMA, thereby generating a self-induced acceleration of the whole process \cite{kolmeder_thermomechanical-chemically_nodate}.

\begin{table}[hb!]
    \centering
    \begin{tabular}{llll}
        \toprule
        \textbf{Authors} & \textbf{Cement} & \textbf{Methods} & \textbf{Characterization type}\\
        & & & \textbf{and conditions} \\
        \midrule
        \citet{boger_clinical_2009} & Vertebroplastic & Oscillatory &Time, 1 Hz, various $T$ \\
        \citet{deusser_rheological_2011} & Vertecem & Oscillatory  & Time, 1 Hz, 23 °C and 37 °C\\
        \citet{dunne_flow_1998} & Unnamed & Capillary & Time\\
        \citet{farrar_rheological_2001} & Various & Oscillatory & Time, 0.05\%, 5 Hz, various $T$ \\
        \citet{kolmeder_characterisation_2013} & Osteopal & Rotational CP  & Time, various $\dot{\gamma}$\\
        & & Capillary & Time, various $\dot{\gamma}$ \\
        \citet{kolmeder_thermophysical_2011} & Unnamed & Oscillatory PP & Time, 1 Hz, 0.1\%, various $T$\\
        \citet{krause_viscosity_1982} & Various & Rotational CP & Time, 0.04 and 1 s$^{-1}$ \\
        & & Capillary & Shear rate, 23 °C \\
        \citet{lepoutre_bone_2019} * & Osteopal & Oscillatory PP  & Time, 50\%, various $f$ and $T$ \\
        & & Injector & Shear rate, various $\dot{\gamma}$ \\
        \citet{lewis_rheological_2002} & Orthoset & Oscillatory CuP & Time, 1 Hz, 18 °C\\
        \citet{lian_biomechanical_2008} * & Simplex & Oscillatory PP & Frequency sweep, 0.2\% \\
        \citet{nicholas_analysis_2007} & Various & Oscillatory PP & Time, 5 Hz\\
       \bottomrule
    \end{tabular}
    \caption{An overview of previous studies found in literature done to characterize PMMA bone cement, the methods used, and the conditions used for measurements. CP and PP stand for cone-plate and parallel-plate setup respectively. Studies marked with * assume the validity of the Cox-Merz rule.}
    \label{tab:literature}
\end{table}

In general, it should be noted that PMMA bone cement is far from behaving like an ideal fluid, rather it has a dough-like consistency while it is curing. The behavior of the bone cement can be categorized as viscoelastic, i.e.~it is a combination of solid-like and fluid-like behavior. Moreover, the behavior is dependent on many factors, including time, flow conditions (e.g.~shear rate), and temperature. Due to this multifactorial dependency, the complete characterization of the bone cement is a challenge. Many previous studies have measured the rheological properties of the bone cements. Different studies use different brands of PMMA bone cement for their measurements. Moreover, the set of conditions used to carry out the measurements is different in each study. Not only the conditions but also the methods used for the measurements are different in each study, e.g.~while some studies report steady shear viscosity from rotational shear measurements, others report complex viscosity from oscillatory measurements. The differences in some of these studies are shown in Table \ref{tab:literature}. Due to these reasons, it is difficult to compare or compile the data from these studies to understand the complete rheological behavior of a bone cement over the range of shear rates and temperatures expected during vertebroplasty. The steady-state and complex viscosities can be compared only using the Cox-Merz rule \cite{cox_correlation_1958}. The Cox-Merz rule is an empirical relationship indicating that for unfilled polymers the shear viscosity dependency on the shear rate can be predicted from the angular frequency dependency of the complex viscosity. There are studies in the literature that show that the Cox-Merz rule is applicable for PMMA \cite{onogi_rheological_1968}. However, the many additives in the bone cement make it a highly filled polymer, and similar materials have been reported to not obey the Cox-Merz rule owing to their high molecular weight and complex intermolecular binding phenomena \cite{rathner_applicability_2021}. Nevertheless, its validity for the PMMA bone cement is assumed in many studies rather than being explicitly investigated. 

Hence, the aim of this work was to understand the flow behavior of the bone cement during injection in its curing phase in conditions relevant to vertebroplasty, taking Vertecem V+ as an example. The main objectives of this work were as follows:
\begin{enumerate}
    \item To quantify various aspects affecting injection flow behavior, namely viscosity, yield stress, loss factor, flow point, etc.~from measurements
    \item To fit the experimental results to mathematical models that describe the rheological behavior, like the power law model, and obtain the model parameters that can be used for prediction and simulation of injection flow behavior
    \item To compare the different methodologies used for rheological measurement, namely injection tests as well as rotational and oscillatory measurements on the rheometer, namely their advantages and limitations
    \item To check the validity of the Cox-Merz rule for the PMMA bone cement
\end{enumerate}

\section{Experimental}
\label{sec:methods}
\subsection{Bone cement}
The bone cement used for this study was purchased as non-sterile bulk material from OSARTIS GmbH (Germany), which is equivalent to the bone cement commercially sold by the name Vertecem\textsuperscript{TM} V+ by DePuy Synthes. The bone cement is supplied as a powder consisting of 40\% zirconium dioxide, 15\% hydroxyapatite, 44.6\% poly(methyl methacrylate) (PMMA), and 0.4\% benzoyl peroxide; and a liquid consisting of 99.35\% methyl methacrylate (MMA) stabilized with 60 ppm hydroquinone, and 0.65\% N,N-dimethyl-p-toluidine (DMPT), making it a highly filled polymer. Typical application of the bone cement uses the provided cement mixing kit (DePuy Synthes), where the initiator and powder are mixed with MMA monomers in 26 grams of powder to 10 mL of liquid ratio, upon which immediately the user is required to push and pull the mixing chamber kit handle from endpoint to endpoint for 20 seconds, at a steady pace of 1–2 strokes per second. The resultant mixture has a dough-like consistency which can be filled into syringes and injected. The polymerization occurs through the free radical polymerization method, wherein the activator DMPT reacts with the initiator benzoyl peroxide to form free radicals. These free radicals react with monomer MMA molecules causing them to form polymer chains of PMMA. PMMA is the main component of bone cement and is responsible for its mechanical properties. The entire polymerization process takes about 20 minutes at room temperature. 

\subsection{Rheometer measurements}

The rheological measurements were performed on an MCR302 rheometer from Anton Paar using a parallel plate (PP) setup. Disposable top and bottom plates were used for the measurements. The diameter of the top plate was 25 mm. The bone cement contains particles of size 5.2 ± 1.5 microns \cite{verrier_evaluation_2012}, hence the gap between the two plates was kept at 1.5 mm to ensure a sufficiently larger gap size compared to the particle size \cite{farrar_rheological_2001} and avoid any particle-scale effects. The temperature during the measurements was controlled using the Peltier module of the rheometer, which maintained the temperature of the bottom plate. The temperature was fixed at 23 °C in all tests unless otherwise specified. The feedback constant temperature control of 23 °C ensures that the exothermic release of heat is not affecting our measured processes. The rheometer and the associated instrumentation were switched on for at least one hour to equilibrate the temperature and all other conditions before carrying out measurements. 

The bone cement prepared using the typically provided mixing kit would produce a large offset of unnecessary material for measurements. On the other hand, the results of the rheological measurements are very sensitive to the mixing conditions. Hence, we established a mixing process for our laboratory conditions that not only produced just the necessary amount of bone cement for each measurement but also gave repeatable results on our benchmark test. In the test, the bone cement sample was subjected to oscillations at a maximum torque of 3 mNm and 1 Hz frequency at 23 °C, with a measuring point every 5 seconds. After multiple trials, the successful and reproducible method consisted of the following steps:
\begin{enumerate}
    \item 2.6 grams of the PMMA powder were weighed in a 10 mL beaker.
    \item Separately, 10 mL of the MMA monomer liquid was prepared in a batch glass vial.
    \item At time $t$ = 0 s, 1.0 mL MMA monomer liquid was dropped into the 10 mL beaker with the powder using a positive displacement pipette, and a stopwatch for measuring the time was started.
    \item The powder and the liquid were stirred using a non-reacting polyetheretherketone (PEEK) stirring road for 20 seconds and counting about 20 rotations.
    \item After 20 seconds, part of the sample was gently dropped onto the rheometer bottom plate and the top plate was gently lowered down to avoid any disturbance. 
    \item Silicone oil was then spread around the sample and the humidity control hood was used to avoid sample evaporation. The test was then started and the time on the stopwatch was recorded. 
\end{enumerate}
The results of the benchmark test are provided in the supporting information in Figure S1. The mixing method produced qualitatively reproducible results with quantitative deviations of up to 25\%. Therefore, for the tests where capturing the bone cement response was sufficient, the measurements were not repeated to conserve material. For tests where the quantitative values were important, we did at least three repetitions of the tests, e.g.~the amplitude and frequency sweep tests which were used to obtain the yield stress and the power law index respectively.

The bone cement used here does not have any waiting time \cite{boger_medium_2011}, so it can be used or injected immediately after mixing. The test on the rheometer commenced once the bone cement was prepared and placed between the parallel plates. The time from the start of mixing to the start of test for all tests varied in the range of 182 $\pm$ 33 seconds. For the characterization of the bone cement, various tests were done using the rheometer, the conditions of which were chosen in the context of the bone cement and vertebroplasty, based on information from previous studies in literature, like \citet{krebs_clinical_2005}, and analytical calculations. For the oscillatory tests, the complex modulus (norm of storage and loss modulus), the loss factor (the ratio of loss to storage modulus), and complex viscosity (norm of real and imaginary viscosity) were of main interest. For the rotational tests, the shear viscosity was the main quantity of interest. These tests are detailed below. 
\begin{itemize}
	\item{\textbf{Test Rh1a}:} In this test, the bone cement was subjected to oscillatory deformation of 0.2\% strain amplitude and 1 Hz frequency for 30 minutes. The test was done to observe the evolution of the bone cement properties with time at a low strain value, i.e.~nearly at rest condition. 
	\item{\textbf{Test Rh1b}:} In this test, the bone cement was subjected to oscillatory deformation of 20\% strain amplitude and 1 Hz frequency for 30 minutes. The test was done to observe the evolution of the bone cement properties in time under large deformations, i.e.~in flow conditions.
	\item{\textbf{Test Rh2a}:} The test applied oscillatory deformation in three stages: (i) 0.1\% strain amplitude, 0.1 Hz (5 mins) (ii) 20\% strain amplitude, 1 Hz (3 mins) (iii) 0.1\% strain amplitude, 0.1 Hz (22 mins). This procedure was done to replicate a typical injection, in which the bone cement is at rest at first after mixing, then it is deformed as it is filled and injected through the syringe and cannula, and finally is at rest after injection. The aim here was to investigate how the bone cement breaks under deformation and recovers thereafter. 
	\item{\textbf{Test Rh2b}:} The test applied oscillatory deformation in three stages: (i) 0.1\% strain amplitude, 0.1 Hz, 23 °C (5 mins) (ii) 20\% strain amplitude, 1 Hz, 23 °C (3 mins) (iii) 0.1\% strain amplitude, 0.1 Hz, 37 °C (22 mins). This procedure was the same as in test Rh2a but the temperature in the last stage was increased to human body temperature to replicate a typical injection into a human vertebra, and thereby investigate the recovery in those conditions.
	\item{\textbf{Test Rh3}:} In this test, rotational shear stress was applied in steps of (i) 0 Pa (ii) 100 Pa (iii) 0 Pa (iv) 500 Pa (v) 0 Pa (vi) 2000 Pa (vii) 0 Pa, of one minute each, to evaluate creep behavior. The shear stress values were chosen \textit{a posteriori} based on the yield stress obtained from the amplitude sweep test Rh5. 
	\item{\textbf{Test Rh4}:} Two tests were done, once where rotational shear strain was applied in steps of (i) 0\% (ii) 1\% (iii) 0\%, of two minutes each, and another in steps of (i) 0\% (ii) 100\% (iii) 0\%, of two minutes each, to evaluate stress relaxation behavior. The strain values were chosen \textit{a posteriori} based on the yield stress obtained from the amplitude sweep test Rh5. 
	\item{\textbf{Test Rh5}:} This was an amplitude sweep test, where the strain amplitude was gradually increased from 0.01\% to 1000\%. The test was done at 1 Hz frequency. The aim of the test was to obtain the yield stress, the crossover point, and the linear and non-linear viscoelastic ranges of the bone cement. 
	\item{\textbf{Test Rh6}:} This was a frequency sweep test, where the frequency was gradually increased from 0.1 Hz to 100 Hz at various strain amplitudes. The strain amplitudes were chosen \textit{a posteriori} as per the different regions obtained from the amplitude sweep test Rh5. The test aimed to obtain the dependence of the complex modulus and the complex viscosity on the oscillation frequency, and to use the complex viscosity-frequency curve for comparison to rotational shear rate sweep (test Rh7a) for checking the validity of the Cox-Merz rule.  
	\item{\textbf{Test Rh7a}:} Various rotational shear rate sweep tests were done to obtain flow curves, i.e.~viscosity change with shear rate, with each test spanning over two to three orders of magnitude of shear rate. In this way, steady-state viscosity values for shear rates from 10$^{-4}$ s$^{-1}$ to 10$^{3}$ s$^{-1}$ were obtained to cover the possible range of shear rate during vertebroplasty. 
	\item{\textbf{Test Rh7b}:} Rotational shear rate sweep from 0.1 to 100 s$^{-1}$ were done with 100-grit sandpaper stuck to the top and bottom plates using double-sided tape. The test was done to check if there was a difference in the result with and without sandpaper, which would provide evidence for the occurrence of wall slip, i.e.~insufficient friction between the rheometer plates and the bone cement sample for proper transfer of applied forces. 
	\item{\textbf{Test Rh8}:} Various tests were done by applying deformation with constant rotational shear rate for 20 minutes. These tests were done at shear rates 10$^{-4}$, 10$^{-3}$, 10$^{-2}$, 10$^{-1}$, 1, and 100 s$^{-1}$, to evaluate the time-evolution of viscosity at flow rates possible during vertebroplasty. 
\end{itemize}

\subsection{Optical microscopy (tests Rh9 and Rh10)}
Two additional tests were done on the rheometer to make bone cement samples for investigation with optical microscopy. In the first test (test Rh9), bone cement was prepared and simply left on the rheometer between the plates while maintaining a 1.5 mm gap for 45 minutes without any action. In the second test (test Rh10), bone cement was prepared and subjected to 15 minutes of rotational 100 s$^{-1}$ shear rate, like in test Rh7b, and then 30 minutes of no action on the rheometer. The two cured bone cement discs formed as a result of these tests were investigated using an Axiotech microscope, which is a material microscope for non-transparent samples, equipped with a 2.5x objective. The images were captured using a digital camera Axiocam 105, and processed using software Axiovision 4.9.1. The microscope, camera, and processing software were provided by Carl Zeiss AG, Germany. Reflected light bright field technique was used for the illumination.

\subsection{Tests on the custom-made injector (Tests Inj1 -- Inj5)}
\begin{figure}[hb!]
    \centering
    \begin{subfigure}{0.75\linewidth}
        \centering
    	\includegraphics[width=\linewidth]{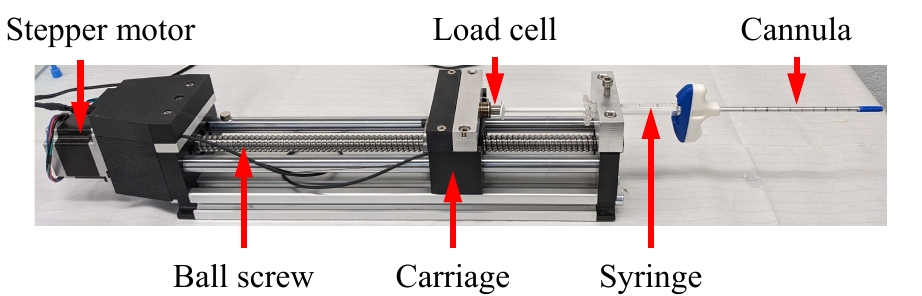}
    	\caption{}
    	\label{subfig:injector}    
    \end{subfigure}
    \begin{subfigure}{0.75\linewidth}
        \centering
        \includegraphics[width=\linewidth]{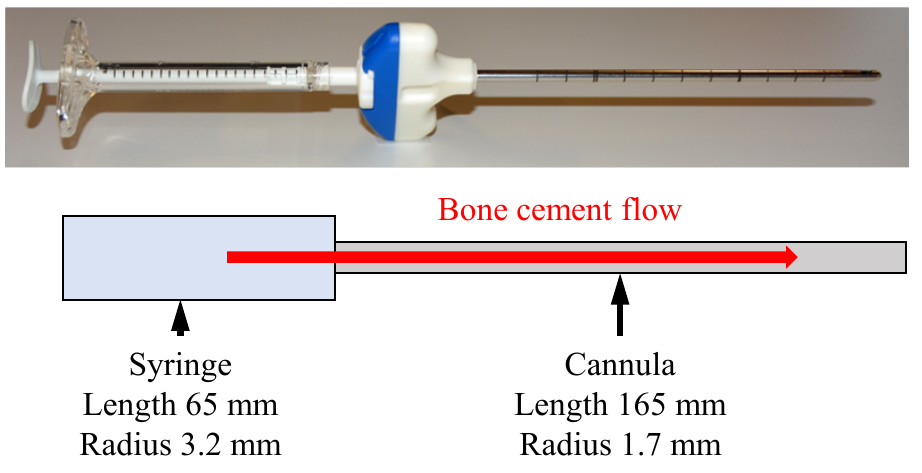}
        \caption{}
        \label{subfig:schematic}
    \end{subfigure}
    \caption{(a) Setup for the injection tests (b) Schematic of the simplified geometry used for analytical calculations}
\end{figure}

The setup for the injection experiments is shown in Figure \ref{subfig:injector}. The injector consisted of a carriage which was driven by a stepper motor using a ball screw with a 5 mm feed per revolution. The stepper motor had a resolution of 1.8° corresponding to 200 steps per revolution, amounting to a calculated carriage resolution of 0.025 mm. The syringe was placed horizontally through a hole at the other end of the device as shown in Figure \ref{subfig:injector} and held in place with a bolt. The cannula was then attached at the end of the syringe. The carriage pushed the plunger at the desired rate which could be programmed using a computer. At the point of pushing the plunger, a 200 N load cell was mounted on the carriage to measure the forces applied to the plunger during injection. 

To prepare the bone cement, a 10 mL syringe was taken and sealed at its open end using a removable plug. The plunger of the syringe was removed and 3 mL of the monomer liquid was poured into it using a pipette. 7.8 grams of the PMMA powder were added directly into the syringe using a funnel and then mixed in the syringe using a PEEK stirring rod for 20 seconds (20 rotations). After mixing the components, the plunger was put back in the syringe and the plug at the other end was removed. The mixed bone cement was then first transferred to a 2 mL syringe until it was full, which was then used to fill up an 8-gauge cannula. The cannula had a side-opening and open front end, of which the side-opening was sealed using a strong adhesive tape so the bone cement could only come out of the front end when injected. The emptied volume of the 2 mL syringe was then again filled up using the remaining bone cement in the 10 mL syringe. At the beginning of the test, the carriage was set at the `unloaded' position, which is far back to be able to mount the filled 2 mL syringe easily. Once the syringe and the cannula were mounted, the carriage was set to `loaded' position, which was just behind the fully pulled back plunger. The `unloaded' and `loaded' positions were programmed in the device so it automatically arrived at them when commanded. At this point, the device was given the command to inject with a specified speed according to the required flow rate. The time from the start of mixing to the start of injection was measured using a stopwatch. The tests, named \textbf{Inj1} to \textbf{Inj5}, were carried out with flow rates of 0.025, 0.05, 0.1, 0.2, and 0.4 mL/s respectively. During the injection, the load cell measured the force required for injection every 0.05 seconds. The time from mixing to the start of injection in all tests was between 230 and 245 seconds.

\subsection{Analytical calculations}
The measurements obtained from the injection tests were then used to obtain the rheological parameters of the bone cement according to the power law constitutive equation 
\begin{equation}
	\mu = K \,\dot{\gamma}^{n-1},
	\label{eq:powerlaw}
\end{equation}
where $\mu$ is the viscosity, $\dot{\gamma}$ is the shear rate, and consistency index $K$ and flow index $n$ are the material parameters. Assuming such a fluid flows through a tube of radius $R$ and length $L$ with flow rate $Q$, the shear stress $\tau$ at the wall relates to the pressure difference $\Delta p$ as
\begin{equation}
	\tau = \frac{R \, \Delta p}{2\, L},
	\label{eq:ana1}
\end{equation}
\begin{equation}
	\Delta p = \frac{2 \, L \, K}{R} \bigg( \frac{1+3\,n}{n} \frac{Q}{\pi\, R^3} \bigg )^{n}
	\label{eq:pres_ana1}	
\end{equation}
Furthermore, the shear rate in the tube can also be obtained using the equation
\begin{equation}
	\dot{\gamma}(r) = \bigg( \frac{1+3\,n}{n} \frac{Q}{\pi \,R^3} \bigg ) . \label{eq:sr_r}
\end{equation}
Detailed derivations of these equations can be found in the works of \cite{lenk_hagen-poiseuille_1978, simpson_newtonian_2009}.

To apply these equations to our problem, the geometry of the apparatus was simplified as shown in Figure \ref{subfig:schematic}. The syringe and the cannula were assumed to be cylinders with the given dimensions. The coupling connecting the two parts had the same inner diameter as the cannula, hence its length was included in the cannula in the simplified geometry. The nozzle of the syringe was short in length and was therefore ignored. The change of cross-section from syringe to cannula would add to the pressure loss in reality, however, this was ignored in our study. We then used the Equation \ref{eq:pres_ana1} to obtain the pressure loss in our system. The pressure loss is then the sum of losses in the syringe $\Delta p_s$ and the cannula $\Delta p_c$, given as
\begin{align}
	\Delta p &= \Delta p_{s} + \Delta p_{c} \nonumber \\
	&= \frac{2 \,L_s \,K}{R_s} \bigg( \frac{1+3\,n}{n} \frac{Q}{\pi\, R_s^3} \bigg )^{n} + \frac{2\, L_c K}{R_c} \bigg( \frac{1+3\,n}{n} \frac{Q}{\pi\, R_c^3} \bigg )^{n} ,
\label{eq:pres_ana2}	
\end{align}
where $L_s = 65$ mm, $R_s=3.2$ mm were the length and radius of the syringe respectively. Similarly, $L_c=165$ mm and $R_c=1.7$ mm were the length and radius of the cannula respectively. Given force equilibrium in the system, the pressure loss $\Delta p$ obtained from Equation \ref{eq:pres_ana2} is the pressure that would be required to be created by the injection force $F$ applied externally on the syringe plunger. Hence, the required injection force $F$ is then given as
\begin{equation}
    F = \Delta p \, (\pi \,R_s^2) .
    \label{eq:force}
\end{equation}

\section{Results and Discussion}

\subsection{Oscillations at constant strain and frequency (Tests Rh1a and Rh1b)}
\begin{figure*}[htb!]
	\centering
	\begin{subfigure}{0.475\linewidth}
		\centering
		\includegraphics[width=\linewidth]{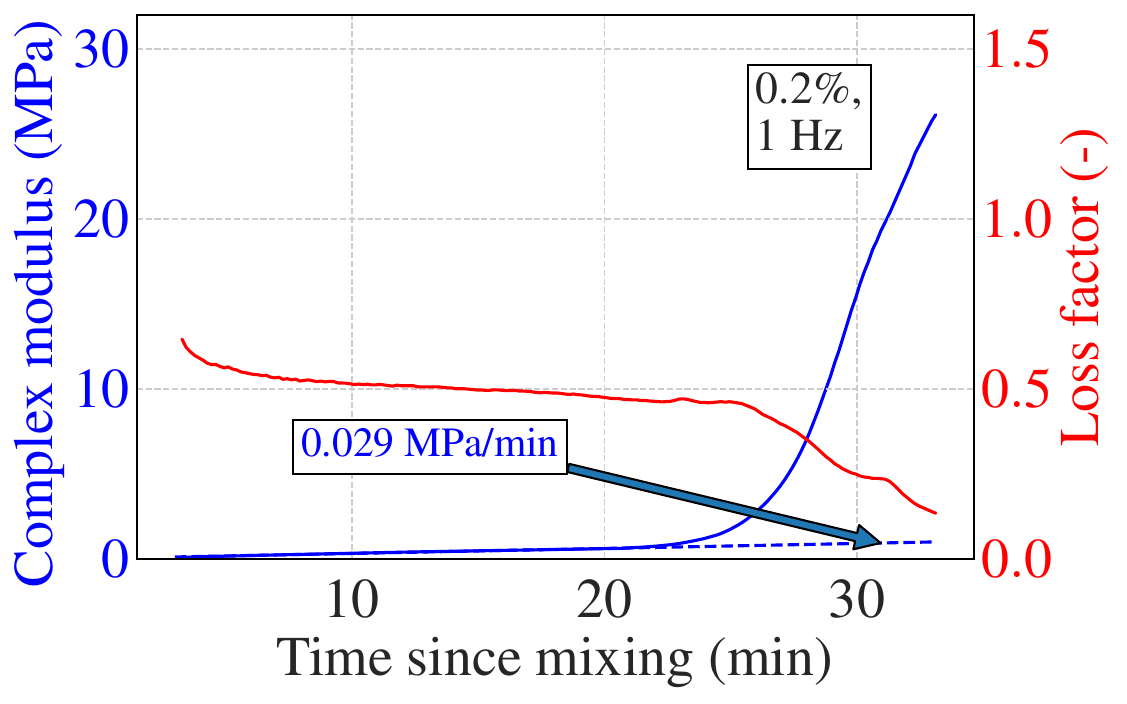}
		\caption{}
		\label{subfig:osc_low}
	\end{subfigure}
	\begin{subfigure}{0.485\linewidth}
		\centering
		\includegraphics[width=\linewidth]{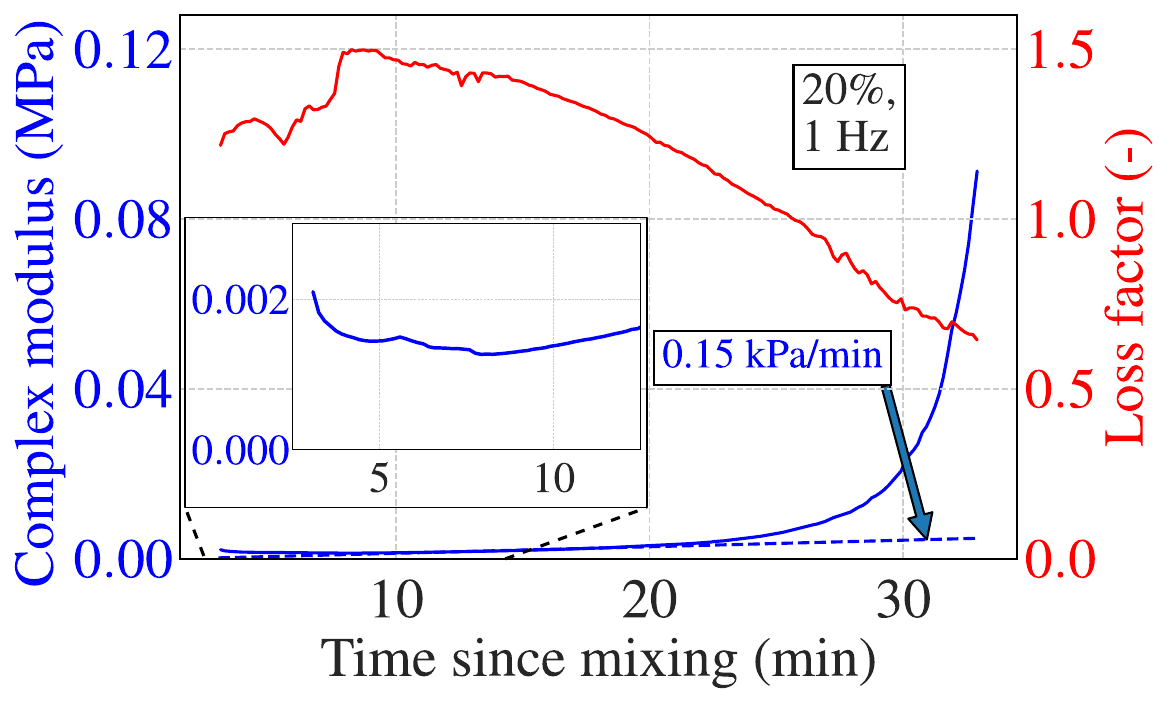}
		\caption{}
		\label{subfig:osc_high}
	\end{subfigure}
	\begin{subfigure}{0.48\linewidth}
		\centering	
		\includegraphics[width=\linewidth]{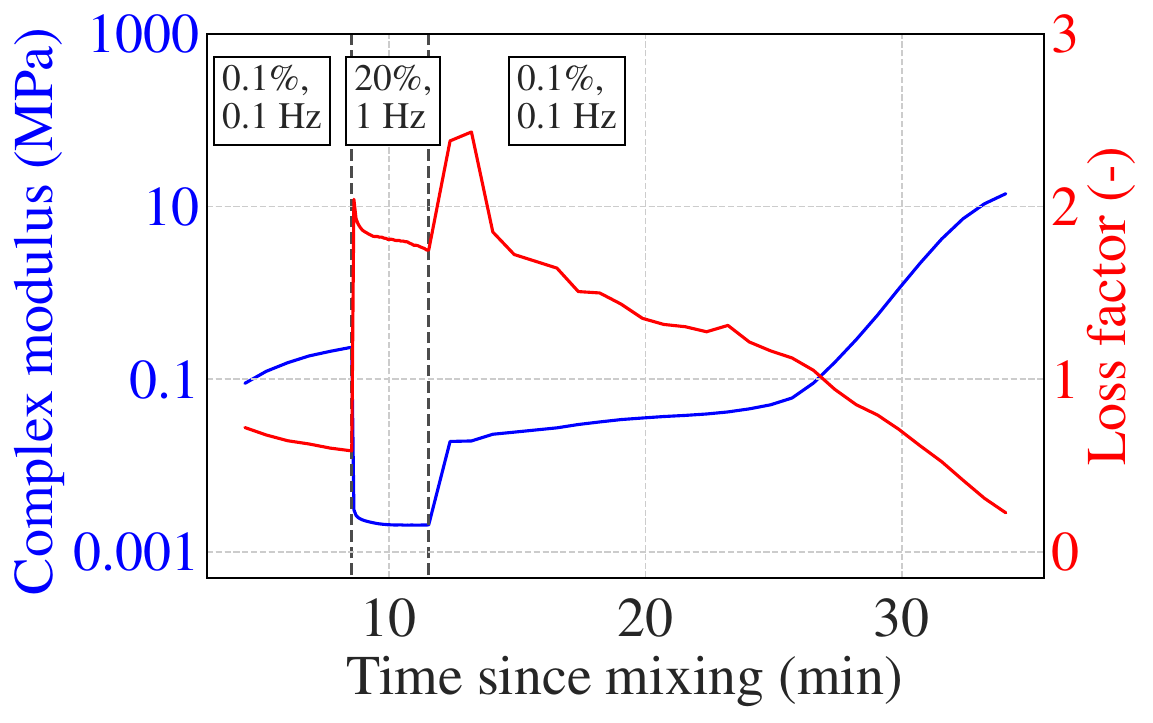}
		\caption{}
		\label{subfig:osc_injsim}
	\end{subfigure}
	\begin{subfigure}{0.48\linewidth}
		\centering	
		\includegraphics[width=\linewidth]{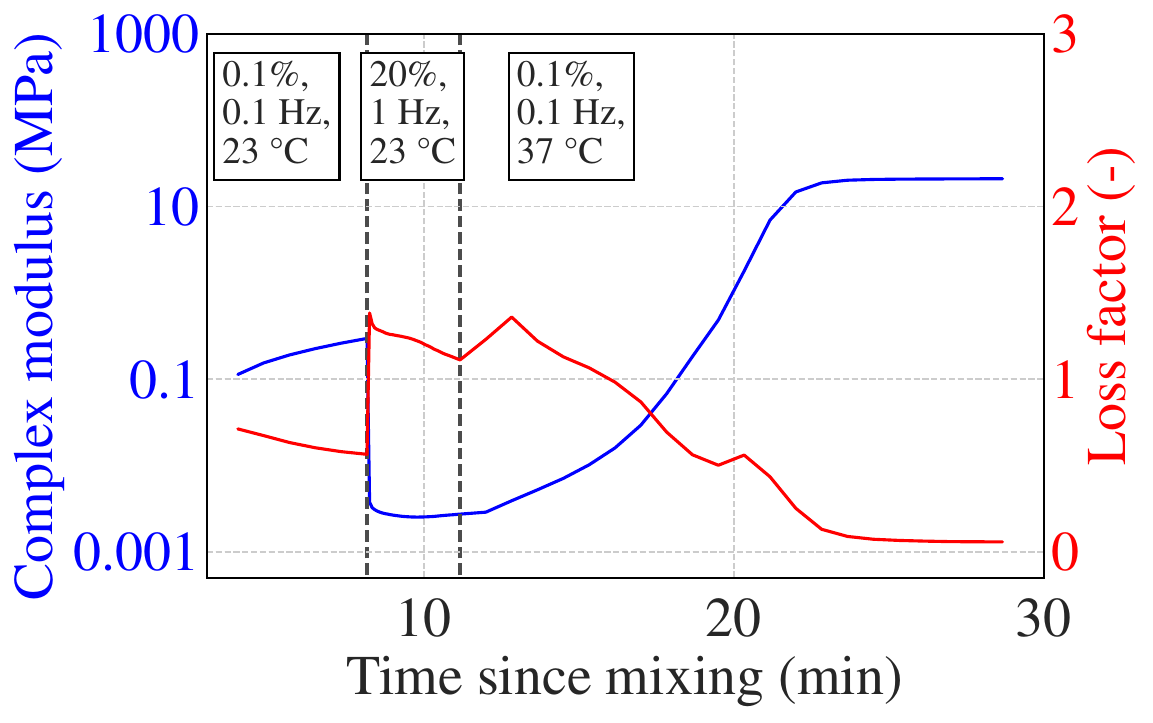}
		\caption{}
		\label{subfig:osc_injsim_temp}
	\end{subfigure}
\caption{Evolution of complex modulus (solid blue line, left Y-axis) and loss factor (solid red line, right Y-axis) as a function of time in oscillatory tests. (a) Test Rh1a: complex modulus increases in two phases indicating the two-phase curing process, the dashed line shows linear regression for complex modulus in the first 20 minutes. (b) Test Rh1b: complex modulus reduces initially as shown in the zoomed part, then resumes increasing in two phases. The dashed line shows linear regression for complex modulus from 10 to 20 minutes. (c) Test Rh2a: A drop in modulus was observed upon increasing strain amplitude, restoring the original amplitude resulted in recovery of modulus and resumption of the two-phase curing. (d) Test Rh2b: similar to Rh2a, but the rapid curing phase was triggered as the temperature was increased to physiological body temperature.}
\label{fig:osc_test}
\end{figure*}

At 0.2\% strain amplitude and 23 °C, as shown in Figure \ref{subfig:osc_low}, the complex modulus of the bone cement increased with time due to the ongoing polymerization. The polymerization occurred with a slow first phase, when the complex modulus increased gradually at a rate of about 29 kPa/min. This is the working phase when the bone cement gradually hardens with time but maintains an injectable paste-like consistency. The slow first phase was followed by a rapid curing phase after about 20 minutes when the bone cement polymerized rapidly, and became solid in the next 5-10 minutes. The complex modulus after solidification was above 25 MPa. The loss factor was about 0.5 - 0.7 until 25 minutes and rapidly decreased thereafter. Hence, the loss factor always remained below 1, implying that the bone cement was predominantly elastic, i.e.~behaving more like an elastic solid than a liquid, even at the early stages of curing. 

When the same test was carried out at the higher 20\% strain amplitude, the complex modulus was about two orders of magnitude lower, as shown in Figure \ref{subfig:osc_high}. Zooming in on the graph showed that initially, the modulus decreased with time. The decrease was an exponential decay at first and a continued decrease afterwards. The exponential decay is typical of viscoelastic stress relaxation behavior, while the continued decrease afterwards could be probably because of fatigue resulting from repeated oscillations. After about 8 minutes, the modulus started to increase again, which was probably due to new polymer chains forming faster than the rate of breaking. The increase was, however, much slower compared to that observed in the test with 0.2\% strain. Interestingly, the loss factor here was greater than 1, implying that the bone cement behaves predominantly like a fluid when it is being deformed, unlike when it is at rest. The loss factor did not go lower than 1 until after 25 minutes, which was deep in the rapid curing phase. 

These results show that the bone cement behavior is predominantly solid when it is at rest or with very small deformations, while high deformations cause the bone cement to behave more like a fluid. This indicates that the bone cement is suitable for injections, despite this not being apparent from the test Rh1a. Furthermore, the curing process is inhibited by the deformations, as is evident from the much lower magnitude as well as a delayed and lower rate of increase in the complex modulus during the entire duration of the test.

The properties of the bone cement continuously evolve because of the curing reaction. Hence, it is important to evaluate the so-called mutation time, which is defined as the time required for change in a given property by a factor equal to the Euler's number $\mathfrak{e}$ \cite{mours_time-resolved_1994}. Taking the complex modulus as the property of interest, the bone cement was found to have a mutation time of about 180 seconds and 700 seconds at 0.2\% and 20\% strain amplitude respectively. For reliable measurements, the measurement times must be much less than mutation times to avoid the effect of the evolving properties in measurements. This is ensured to be the case for most of our measurements shown later in this work. As an exception, the measurement times are not negligible in the initial few points in the frequency sweep, but they are still much smaller than the mutation times.  

\subsection{Replication of injection (Tests Rh2a and Rh2b)}
Figure \ref{subfig:osc_injsim} shows the results of the test Rh2a. During the initial stage of the test, i.e.~when the strain amplitude and frequency were low, the complex modulus increased gradually due to polymerization. This rate was slightly faster than at the initial time of Rh1a (Figure \ref{subfig:osc_low}) test due to the lower strain used. In the next stage of the test, i.e.~when it was subjected to higher strain and frequency, the modulus immediately dropped to 1\% of the original magnitude and the loss factor jumped to a value higher than 1. This again showed the bone cement's transition to more fluid-like behavior as soon as it is deformed during polymerization. This was followed by an exponential decay of the modulus, as was also observed in the previous test Rh1b (Figure \ref{subfig:osc_high}). When the bone cement is brought to (almost) rest again in the final stage of the test (i.e.~at about 12 minutes), there is an immediate but small recovery of about 8\% in the modulus. The polymerization is resumed and the rapid curing phase occurs at about a similar time of 25 minutes, after which the bone cement solidifies. We note, however, that the properties of bone cement modulated in this way (simulating a more injection-like scenario), are different to those measured at low deformations in test Rh1a (Figure \ref{subfig:osc_low}). For example at the end of the test, the complex modulus is 2.5-fold lower and the loss modulus is about 2-fold higher due to structural disturbances during curing. Carrying out the same test with the bone cement kept at 37 °C in the last stage of the test (Rh2b), there is hardly any recovery, however, the rapid curing phase of the bone cement is triggered immediately. The bone cement is a fully cross-linked polymer solid at 22 minutes, as evident from the plateaued complex modulus and the loss factor value of 0.05. 

Figure \ref{subfig:osc_injsim_temp} shows that the curing time is significantly shortened when the bone cement is exposed to higher temperature. Once inside the vertebra of the patient, the bone cement starts curing rapidly. A similar test was also done by \citet{deusser_rheological_2011}, in which they observed much slower curing, likely due to the torque-controlled nature of their measurement as opposed to our amplitude-controlled measurement. Nevertheless, they also observed a shift in the curing rate when the temperature was increased from 23°C to 37°C. Vertebroplasty is often done in steps of smaller injections, in which case practitioners need to be aware that the bone cement cures rapidly when exposed to the physiological body temperature. If the subsequent injections are not done in a timely manner, the viscosity of the cement inside the vertebra could get much higher than the one being injected. This viscosity difference could result in disadvantages like requiring higher injection force, unintuitive injection force development throughout the injection, and unstable flow patterns inside the vertebra, as was highlighted in a previous work \cite{trivedi_continuum_2023}.

\subsection{Step stress (Test Rh3) and step strain (Test Rh4)}
\begin{figure*}[htb!]
	\centering
	\begin{subfigure}{0.45\linewidth}
		\centering
		\includegraphics[width=\linewidth]{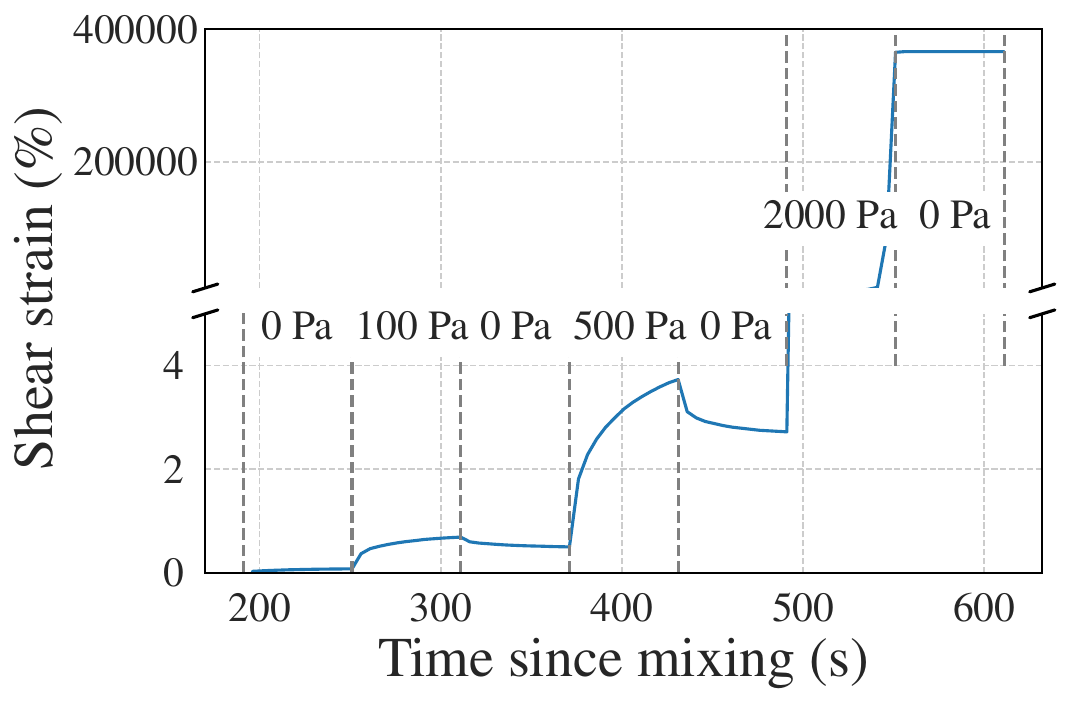}
		\caption{}
		\label{subfig:creep}
	\end{subfigure}
	\begin{subfigure}{0.44\linewidth}
		\centering
		\includegraphics[width=\linewidth]{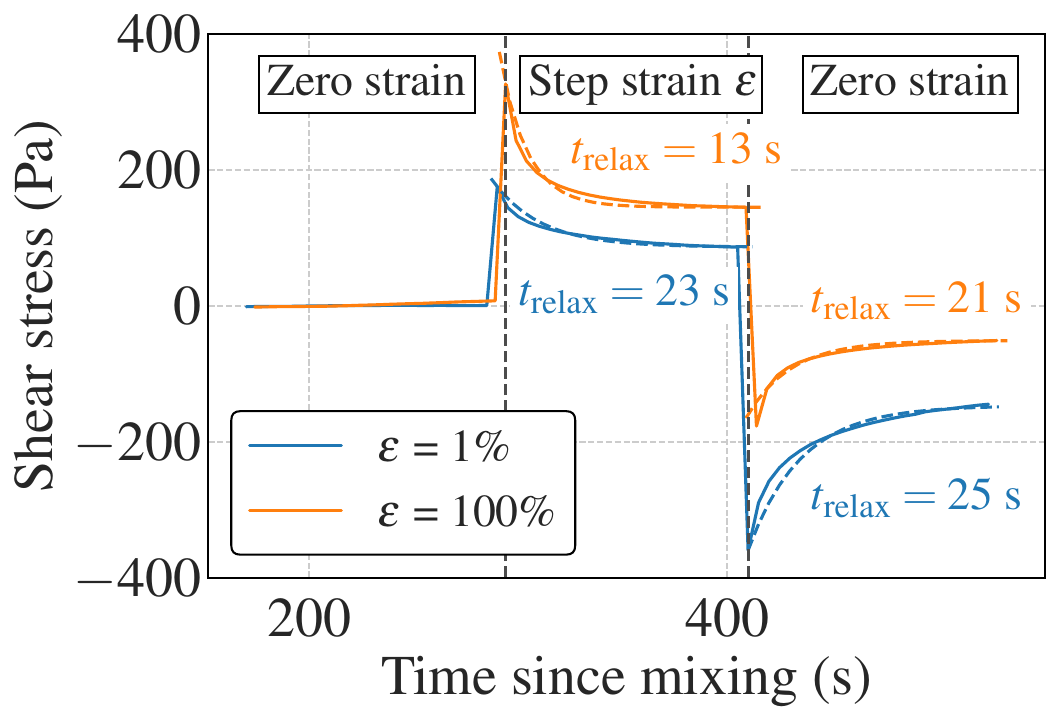}
		\caption{}
		\label{subfig:stress_relaxation}
	\end{subfigure}
	\begin{subfigure}{0.49\linewidth}
		\centering
		\includegraphics[width=\linewidth]{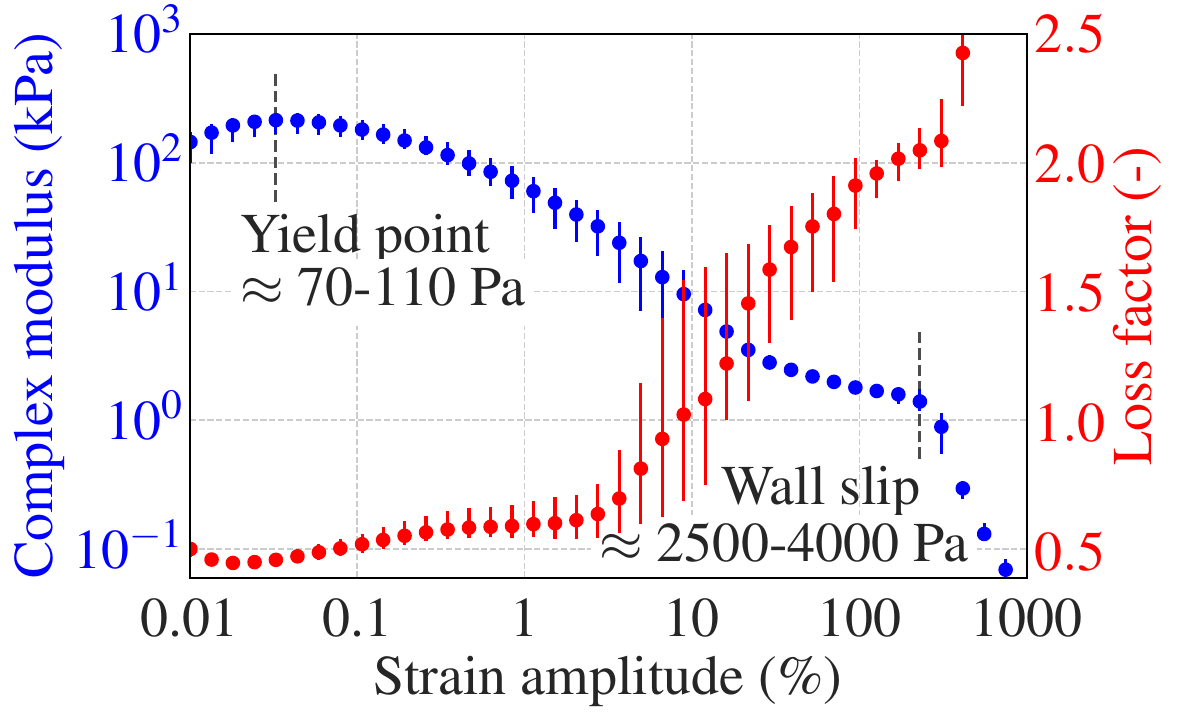}
		\caption{}
		\label{subfig:ampsweep1}
	\end{subfigure}
	\begin{subfigure}{0.45\linewidth}
		\centering
		\includegraphics[width=\linewidth]{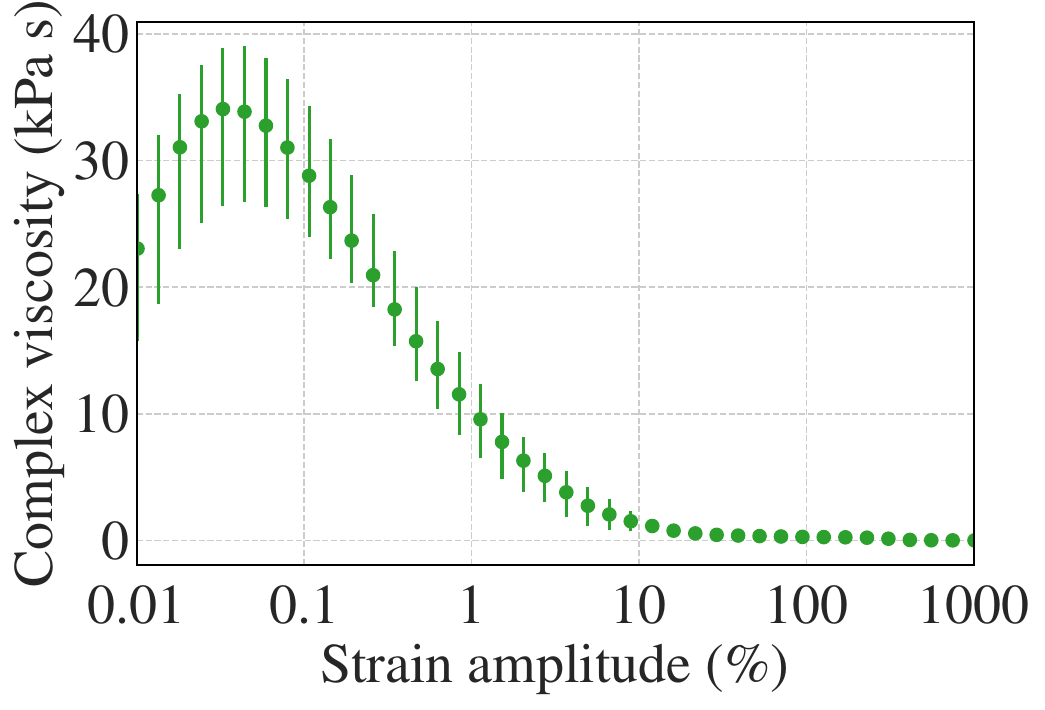}
		\caption{}
		\label{subfig:ampsweep2}
	\end{subfigure}
	\begin{subfigure}{0.49\linewidth}
		\centering
		\includegraphics[width=\linewidth]{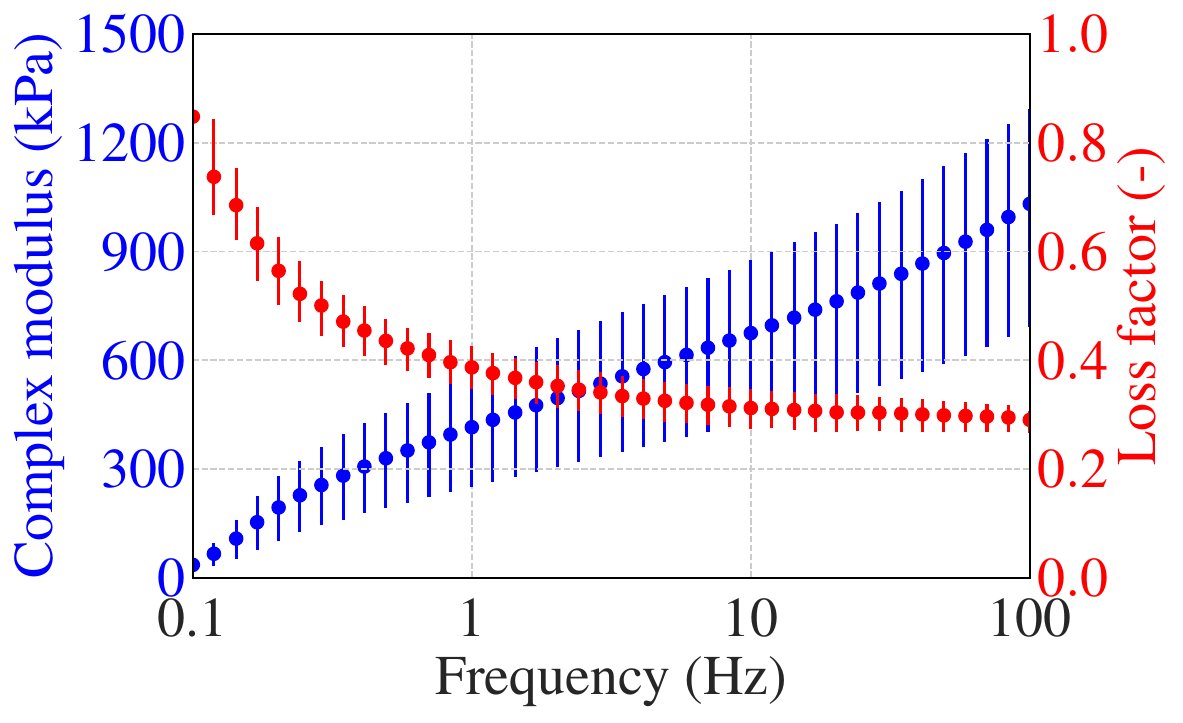}
		\caption{}
		\label{subfig:freqsweep1}
	\end{subfigure}
	\begin{subfigure}{0.45\linewidth}
		\centering
		\includegraphics[width=\linewidth]{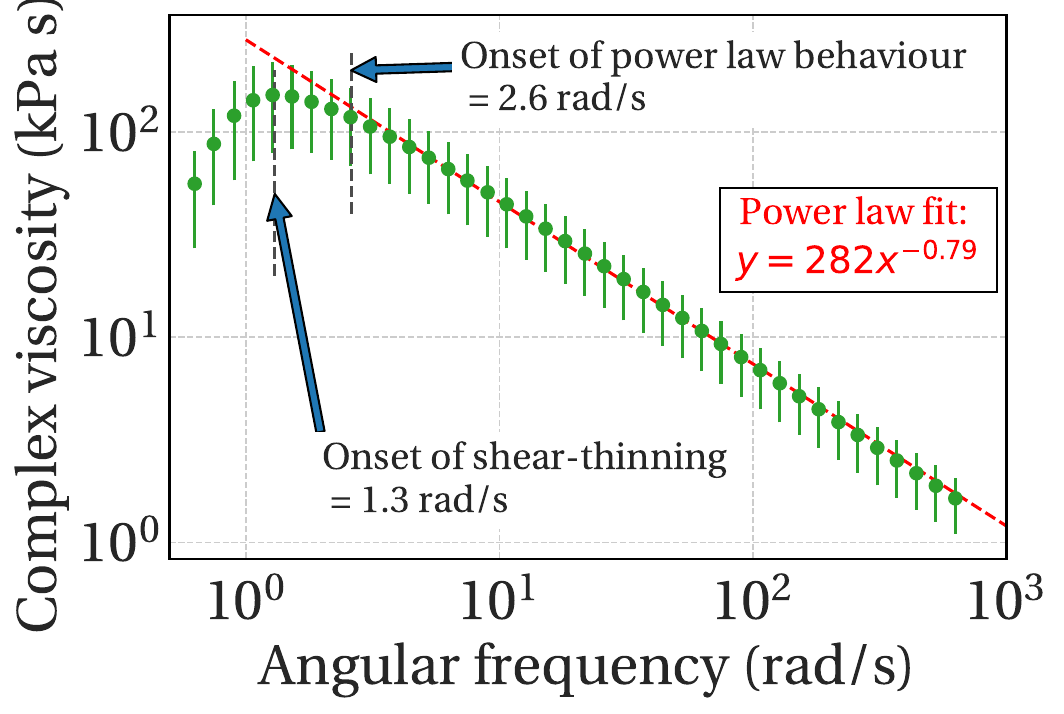}
		\caption{}
		\label{subfig:freqsweep2}
	\end{subfigure}
	\caption{(a) Test Rh3: Creep behavior of bone cement upon applying stress in consecutive steps. The steps are separated by dashed gray lines. (b) Test Rh4: Stress relaxation behavior of bone cement upon applying step strain. The steps are separated by dashed gray lines. The shaded region in the following graphs indicates the range of values and the dashed line inside the region indicates the average. (c) Test Rh5: Complex modulus (blue) and loss factor (red) during amplitude sweep. (d) Test Rh5: Complex viscosity (green) during amplitude sweep (e) Test Rh6: Complex modulus (blue) and loss factor (red) during frequency sweep at 0.01\% strain (f) Test Rh6: Complex viscosity (green) during frequency sweep at 0.01\% strain.}
	\label{fig:amp_freq_sweep}
\end{figure*}
Figure \ref{subfig:creep} shows the behavior of the bone cement when subjected to step stress. The resultant strain was observed to be exponentially decaying, which is typical of viscoelastic behavior, both when stress was applied and then removed. Upon removal of the stress, the strain did not converge back to zero, which indicated that there was permanent deformation in the sample. At very high stress, i.e.~at 2000 Pa, the bone cement showed almost purely viscous behavior as it deformed continuously, and there was no recovery after removal of the stress. 

Figure \ref{subfig:stress_relaxation} shows the stress response of the bone cement to step strain. Two tests were carried out with 1\% and 100\% step strain respectively. The resultant stress was maximum at the start of the applied step strain, and then decreased exponentially with time to a non-zero stress value, typical of viscoelastic stress relaxation behavior. As the applied strain was removed, returning to the zero strain state required stress in the opposite direction, which again showed similar relaxation behavior. The decay curve could be fit to the relaxation equation 
\begin{equation}
	\sigma(t) - \sigma_\infty = (\sigma_0 - \sigma_\infty) e^{-t/t_\text{relax}},
\end{equation}
where $t$ is the time since the start of step strain, $\sigma(t)$ is the stress at time $t$, $\sigma_0$ is the stress at the start of relaxation, $\sigma_\infty$ is the final stress the material relaxes to, and $t_\text{relax}$ is the relaxation time. The relaxation times obtained from curve fitting are shown in Figure \ref{subfig:stress_relaxation}. Compared to the case of 1\% strain, the relaxation times in the case of 100\% step strain are lower, indicating that the bone cement relaxes faster owing to lesser resistance to deformation. This reinforces our previous observation that the bone cement has a more fluid-like behavior while it is being deformed. 

\subsection{Amplitude sweep (Test Rh5)}
Figures \ref{subfig:ampsweep1} and \ref{subfig:ampsweep2} show the range and mean of the complex modulus, loss factor, and complex viscosity for three amplitude sweep tests on the bone cement. The modulus grew initially, likely because the strain was too small to inhibit the polymerization. At 0.02 - 0.04\% strain, the modulus stopped growing and started lowering afterwards. There are various definitions of the yield stress used in the literature. In this work, we call the stress at which the complex modulus and the complex viscosity start to lower as the yield stress, which was observed to be around 70-110 Pa for this bone cement. This is the point at which the ability of the bone cement to resist deformation starts to lower, and the bone cement begins to flow. Hence, this parameter is important to understand the injectability of the bone cement. The given yield stress range corresponds to injection force $F$ of about 0.83 N according to the equation $F = \pi R_s^2\, (\frac{2 L_c}{R_c} + \frac{2 L_s}{R_s})\,\sigma_y$ obtained from Equation \ref{eq:ana1}, which is relatively small and hence corroborates the use of bone cement in clinical injections. However, it is important to note that this value holds only for 295 $\pm$ 20 seconds from the time of mixing, and would increase the more it is allowed to rest. Another interesting point is the crossover point, or the flow point, which is where the loss factor goes above 1, i.e.~the bone cement transits to more fluid-like than solid-like behavior. This occurs in the 1-10\% strain amplitude range or 500-1000 Pa stress ($\approx$ 3-7 N of injection force). Finally, there was a sudden drop in the modulus near 200\% strain or 2500-4000 Pa stress mark, which was likely because of insufficient friction between the rheometer plates and the bone cement, causing the plates to slip. The evidence for this phenomenon, also referred to as wall slip, is provided in later sections. 

\subsection{Frequency sweep (Test Rh6)}
Figures \ref{subfig:freqsweep1} and \ref{subfig:freqsweep2} show the range and mean of the complex modulus, loss factor, and the complex viscosity over three frequency sweep tests done at 0.01\% strain amplitude, i.e.~within the linear viscoelastic limit as seen from the amplitude sweep test in Figure \ref{subfig:ampsweep1}. We observed that the complex modulus kept increasing throughout the test. On the other hand, the complex viscosity initially increased slightly, then kept dropping as the angular frequency increased beyond 1.3 rad/s. This angular frequency can be considered the inception point of shear-thinning behavior. The shear-thinning of the viscosity according to power law, i.e.~Equation \ref{eq:powerlaw}, started at about 2.6 rad/s. Fitting the curve to the power law equation yields the power law flow index $n=0.21$. A value of $n$ between 0 and 1 indicates shear-thinning, which is expected for the PMMA bone cement. Measurements at higher angular frequencies could not be executed due to the rheometer limitation of a maximum frequency of 100 Hz. The results of this test at other strain amplitudes are shown in upcoming sections for checking the validity of the Cox-Merz rule.

\subsection{Shear rate dependence (Tests Rh7a, Rh7b, and Rh8)}
\begin{figure*}[htb!]
	\centering
	\begin{subfigure}{0.46\linewidth}
		\centering
		\includegraphics[width=\linewidth]{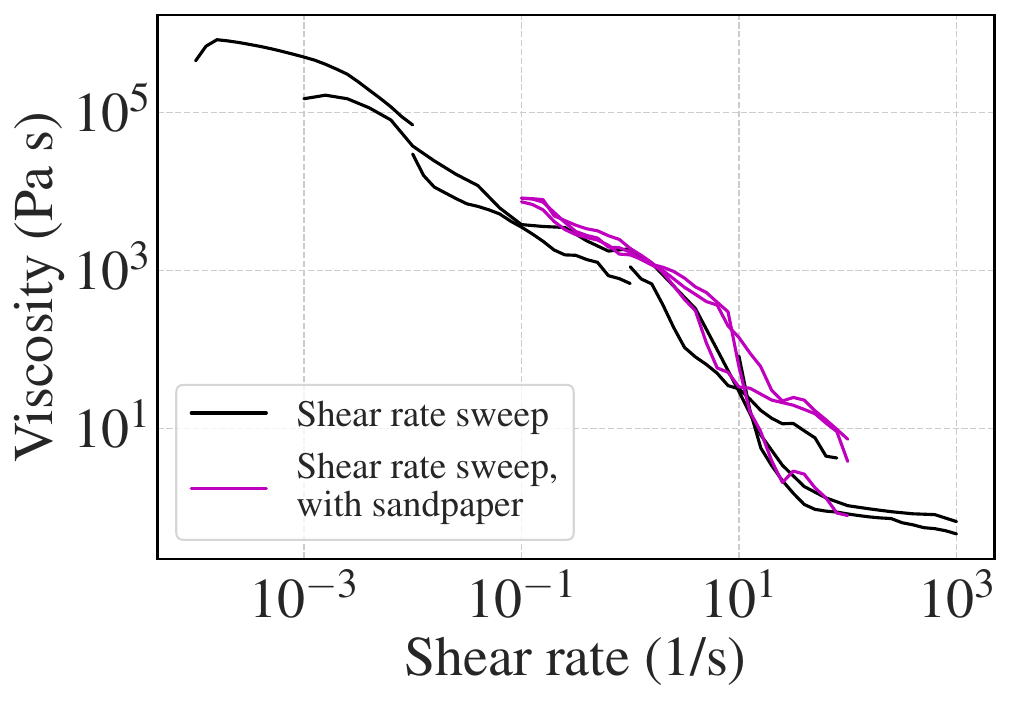}
		\caption{}
		\label{subfig:visc_sr}
	\end{subfigure}
	\begin{subfigure}{0.48\linewidth}
		\centering
		\includegraphics[width=\linewidth]{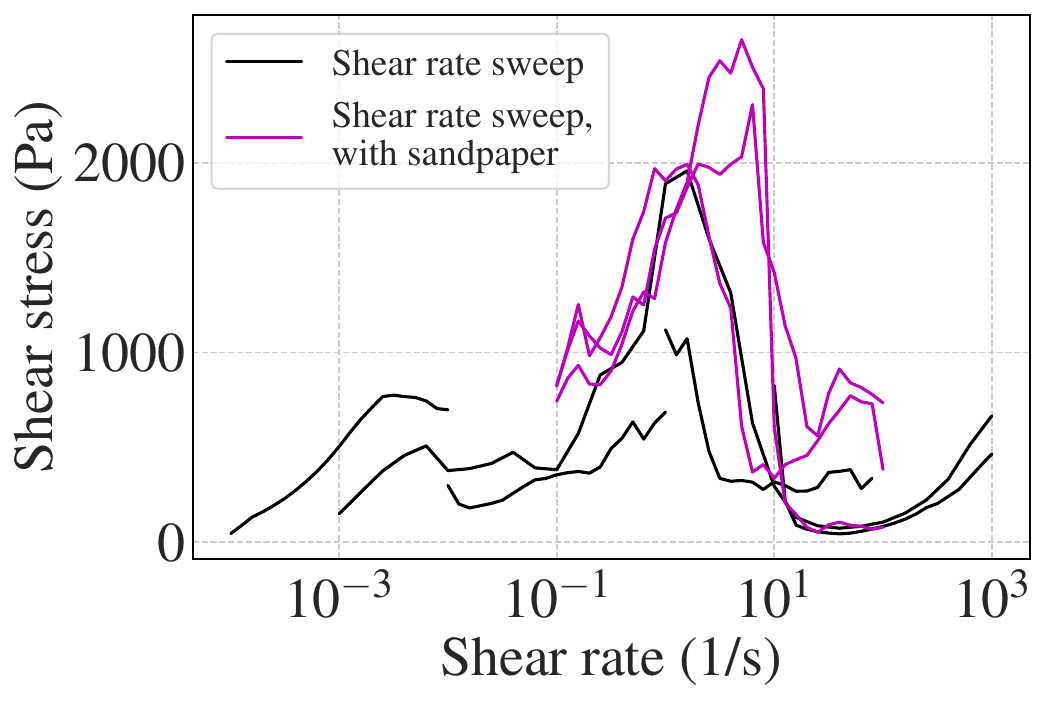}
		\caption{}
		\label{subfig:ss_sr}
	\end{subfigure}
	\begin{subfigure}{0.95\linewidth}
		\centering
        \includegraphics[width=\linewidth]{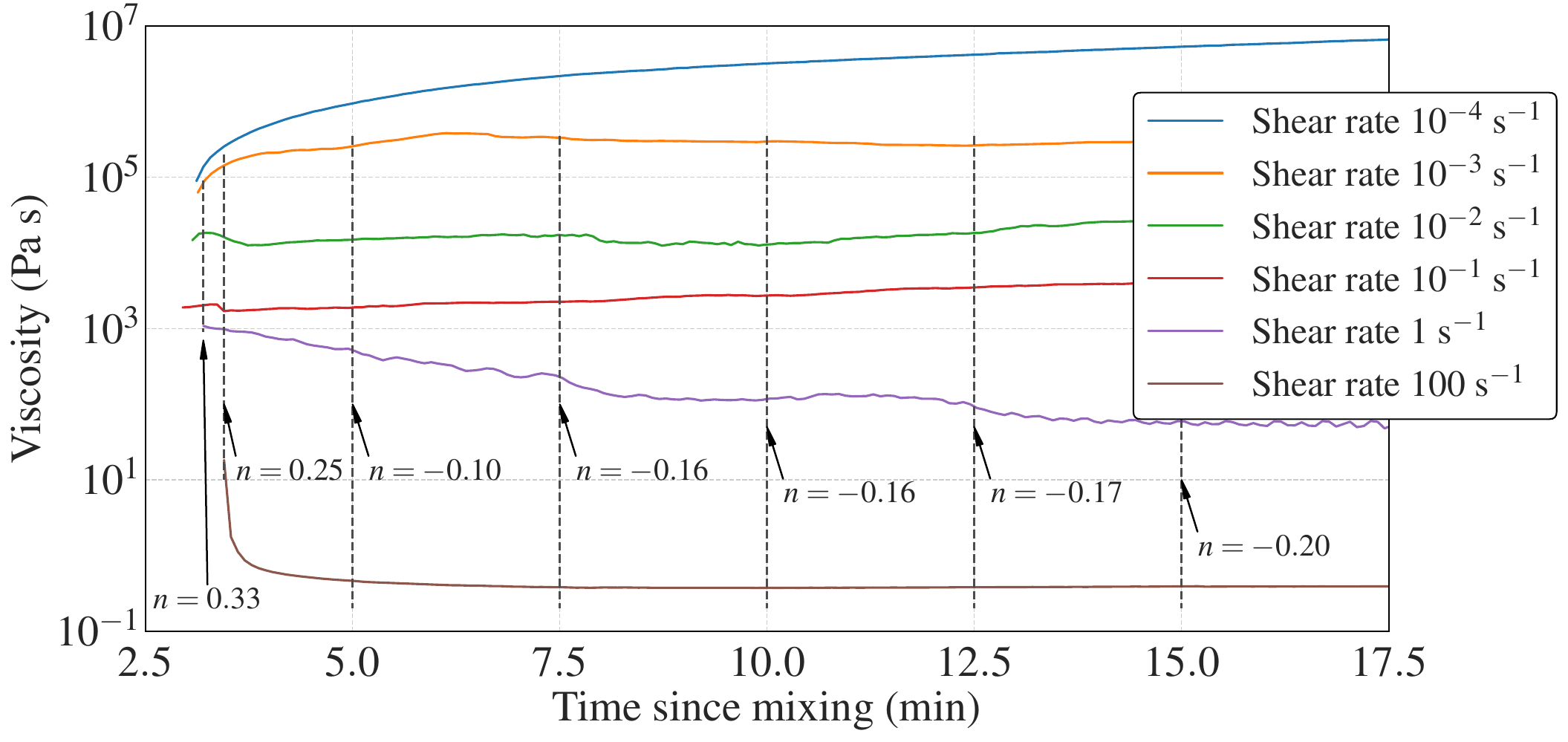}		
		\caption{}
		\label{subfig:visc_time}
	\end{subfigure}
\caption{(a) Flow curves from tests Rh6 (dashed red lines), Rh7a (solid black lines) and Rh7b (dashed black lines) (b) Shear stress plotted against shear rate from tests Rh7a (solid black lines) and Rh7b (dashed black lines) (c) Viscosity evolution as a function of time at different shear rates from tests Rh8. Values of $n$ are obtained by fitting the viscosity values at the instances indicated by dashed lines to the power law equation (Equation \ref{eq:powerlaw}).}
\label{fig:flow_curves}
\end{figure*}

Figure \ref{subfig:visc_sr} shows the results from the various steady shear rate sweep tests performed in rotational mode, also known as flow curves. The measured viscosity was not affected by whether the test was done over smaller shear rate ranges or all the way from 10$^{-3}$ to 10$^3$ s$^{-1}$. This implied that the duration of the test did not affect the results, at least within the first 10 minutes of mixing. The influence of curing in the form of increasing viscosity was seen at very low shear rates in the order of 10$^{-4}$ s$^{-1}$. The shear-thinning is steady until about 1 s$^{-1}$. Upon curve-fitting, the flow index $n$ was obtained in the range of 0.20-0.30, and the consistency index $K$ in the range of 700-1100 Pa s$^n$. 

As the shear rate went above 1 s$^{-1}$, the viscosity drop was noticeably steeper, until it started plateauing at shear rates greater than 20 s$^{-1}$. The values of $n$ were less than zero in this range, implying that the shear stress reduced as the shear rate was increased, as seen in Figure \ref{subfig:ss_sr}. This could be a result of the structural breakdown of the material or a measurement artifact due to the suspected wall slip effect. Performing the same tests using sandpaper attached to both the top and bottom plates (see test Rh7b in `Experimental' section), the results showed a slight improvement in the maximum shear stress achieved (Figures \ref{subfig:visc_sr} and \ref{subfig:ss_sr}). This provided first evidence that the drop in shear stress indeed occurred due to wall slip. This was confirmed from visual inspection through microscopy, as detailed in a later section.

Figure \ref{subfig:visc_time} shows results of tests Rh8, i.e.~time evolution at various constant shear rates. At the lowest shear rate 10$^{-4}$ s$^{-1}$, the bone cement showed a linear increase in viscosity with time of about 8 (kPa s)/s. At lower shear rates, i.e.~at 10$^{-3}$ s$^{-1}$, 10$^{-2}$ s$^{-1}$, and 10$^{-1}$ s$^{-1}$,  lower magnitudes and small dips in the viscosity can be seen, indicating retardation in the curing of the bone cement. This has also been observed for inorganic bone cements by \citet{sahin_preshearing_2020}.  At 1 s$^{-1}$ shear rate, the viscosity kept decreasing for almost the entire duration of the test. At 100 s$^{-1}$ shear rate, the viscosity dropped almost entirely in the early stages (first 30 seconds) of the test. Viscosity values taken at specific points in time for each of these tests were used to calculate the flow index $n$ with time. For this, the viscosity for shear rate 10$^{-4}$ s$^{-1}$ was excluded since previous tests showed it was too low to induce shear-thinning. The results showed that the flow index $n$ was 0.33 in the beginning and 0.25 if the initial value from 100 s$^{-1}$ is taken into calculation. These values are in the same range as observed in tests Rh7a for shear rate less than 1 s$^{-1}$. However, the value of $n$ kept dropping with time and became negative soon after the initial stages, as seen at the 5-minute mark in Figure \ref{subfig:visc_time}. In fact, this was true even when the viscosity values from shear rates 1 and 100 s$^{-1}$ were excluded from the calculation for $n$. This implied that the viscosity values only at the beginning of the test gave plausible shear-thinning characteristics. The same could be said for the consistency index, which when interpreted as the viscosity at unit shear rate, yields $K = 1083$ Pa s$^n$ in the beginning and then decreases with time. Note that the power law model gave similar fitting as the Herschel-Bulkley model, which is generally used for materials with yield stress. The comparison is provided in the supporting information in Figure S2 and Figure S3. 

\subsection{Validity of the Cox-Merz rule for PMMA bone cement}
\begin{figure}[htb!]
    \centering
    \includegraphics[width=0.9\linewidth]{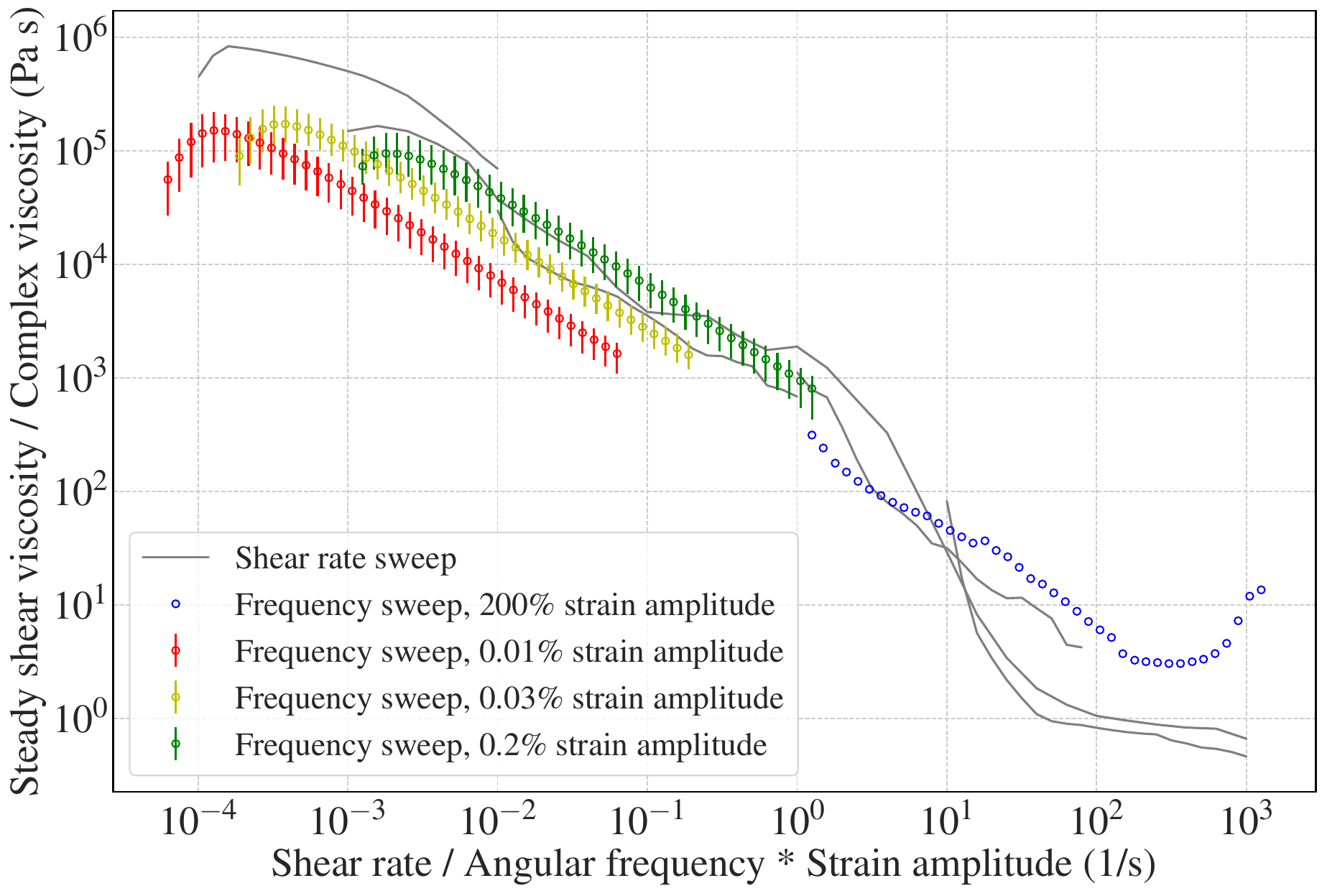}
    \caption{Validation of the Cox-Merz rule}
    \label{fig:cox-merz}
\end{figure}
According to the Cox-Merz rule, for the same magnitude of shear rate and angular frequency, the respectively measured steady shear viscosity and complex viscosity must also be equal \cite{cox_correlation_1958}. To extend this rule to materials with yield stress, \citet{doraiswamy_coxmerz_1991} suggested multiplying the angular frequency by the strain amplitude for the Cox-Merz rule to be true. To validate this for the Vertecem V+ bone cement, the results from the shear rate sweep and three repetitions of frequency sweep measurements at various strain amplitudes are superimposed in Figure \ref{fig:cox-merz}. Note that, as we learned from the amplitude sweep measurements shown in Figure \ref{subfig:ampsweep1}, the 0.01\% strain amplitude is within the linear viscoelastic region, 0.03\% is near the transition to non-linear region, 0.2\% is in the non-linear region. The frequency sweep curves for 0.01\% did not coincide with those of steady shear viscosity, but the same were closer for 0.03\%. The graphs coincided at 0.2\%, i.e.~when the applied strain amplitude was in the non-linear viscoelastic range. As a check, a single frequency sweep was carried out at 200\% strain amplitude, which is beyond the crossover point. The resulting curve seemed to extrapolate the curves from the tests with 0.2\% strain, and coincided with the rotational shear rate graph until the aforementioned dip in viscosity due to wall slip. This is surprising, since conventionally, the frequency sweep is carried out at a strain amplitude lying within the linear viscoelastic range. However, our measurements show that the Cox-Merz rule is indeed valid, but only in the non-linear viscoelastic regime. A possible explanation is provided by \citet{shafiei-sabet_rheology_2012}, which says that the PMMA structures formed as a result of polymerization break under rotational shear flow, but the deformations during the oscillatory tests within the linear viscoelastic region are too low to affect these structures. Hence, the curves of the rotational tests and the oscillatory tests do not coincide in the linear viscoelastic region, but as the strain amplitude goes higher in the non-linear regime, the polymer structures break like in the case of rotational shear flow, and the curves coincide. This reveals that the validity of the Cox-Merz rule for the Vertecem bone cement is conditional, i.e.~only in the non-linear viscoelastic regime. 

\subsection{Optical microscopy and visual inspection}
\begin{figure*}[htb!]
	\centering
	\includegraphics[width=\linewidth]{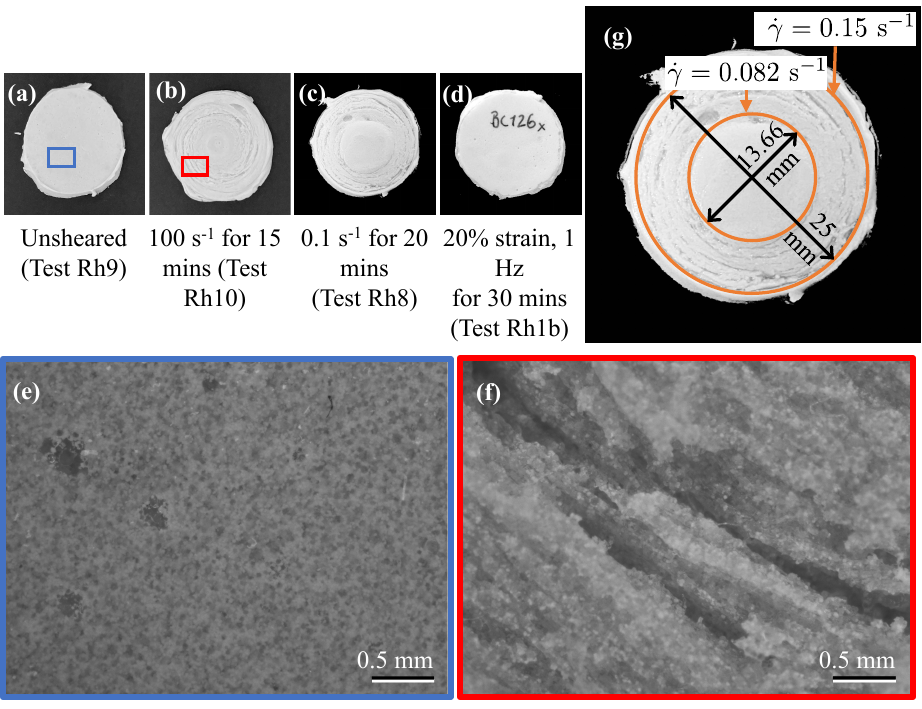}
	\caption{Bone cement samples of 25 mm diameter and 1.5 mm thickness after undergoing testing and curing on the parallel plate rheometer setup are shown in (a), (b), (c), and (d). Microscopic images taken using 2.5x objective for samples in (a) and (b) are shown in (e) and (f) respectively, giving a closer look at the ``ridge"-like formations on the sheared sample. Geometric analysis of the sample in (c), shown in (g), reveals the radius and shear rate beyond which the ridges were formed.}
	\label{fig:photos}
\end{figure*}
Examples of cured bone cement samples after tests on the rheometer are shown in Figure \ref{fig:photos}. These samples were inspected microscopically to gain better insight into the results. Figure \ref{fig:photos}a shows an unsheared sample simply left at rest between the rheometer plates until it was cured, from test Rh9. Figure \ref{fig:photos}b shows a sample subjected to 100 s$^{-1}$ for 15 minutes before letting it cure, from test Rh10. We observed circular ``ridge''-like formations formed in the sheared sample as a result of continuous rotational deformation. The images from optical microscopy, shown in Figure \ref{fig:photos}e and \ref{fig:photos}f, showed that the depth of the ridges went almost through the entire thickness of the sample. The same ridges were observed even in the sample sheared at 0.1 s$^{-1}$ for 20 minutes (referred to test Rh8, Figure \ref{subfig:visc_time}), as shown in Figure \ref{fig:photos}c. A closer look, shown in Figure \ref{fig:photos}g, revealed that the sample looked normal in the center up to a radius of about 13.66 mm, and the ridges were formed only beyond this radius. In a parallel plate rheometer, the shear rate is not uniform across the sample, hence the specified shear rate is actually the average shear rate along the radius. Given that the shear rate at the periphery is 1.5 times the specified shear rate, and the shear rate increases linearly with the radius, we could calculate the critical shear rate as 0.082 s$^{-1}$, beyond which the ridge formation was observed. 

Interestingly, there were no visible ridges in the sample exposed to oscillatory deformation with 20\% strain and 1 Hz for 30 minutes (refer test Rh1b, Figure \ref{subfig:osc_high}), as shown in Figure \ref{fig:photos}d. No net deformation and slip occurs because of the reversible nature of the oscillatory torsion. On the other hand, rotational shearing on the rheometer causes deformations in the bone cement that result in the formation of deep circular ridges, likely stemming from the combination of elastic-like instabilities and possibly Taylor-type vortex flows, as has been observed for other types of materials \cite{byars_spiral_1994, wychowaniec_elastic_2021, shaqfeh_purely_1996}. This could happen even at low shear rates if the sample is sheared for a sufficiently long time. The ridges would not only cause a loss of contact with the top plate, but also a lack of sufficient contact between adjacent flow layers of the bone cement needed for viscous resistance, which would result in underestimation of the viscosity, as has been observed in our rotational rheometer measurements. Hence, this phenomenon limits the conditions in which the rotational rheometry can be done on the bone cement.

\subsection{Tests on the custom-made injector (Tests Inj1 -- Inj5)}
\begin{figure*}[htb]
	\begin{subfigure}{0.48\linewidth}
		\centering
		\includegraphics[width=\linewidth]{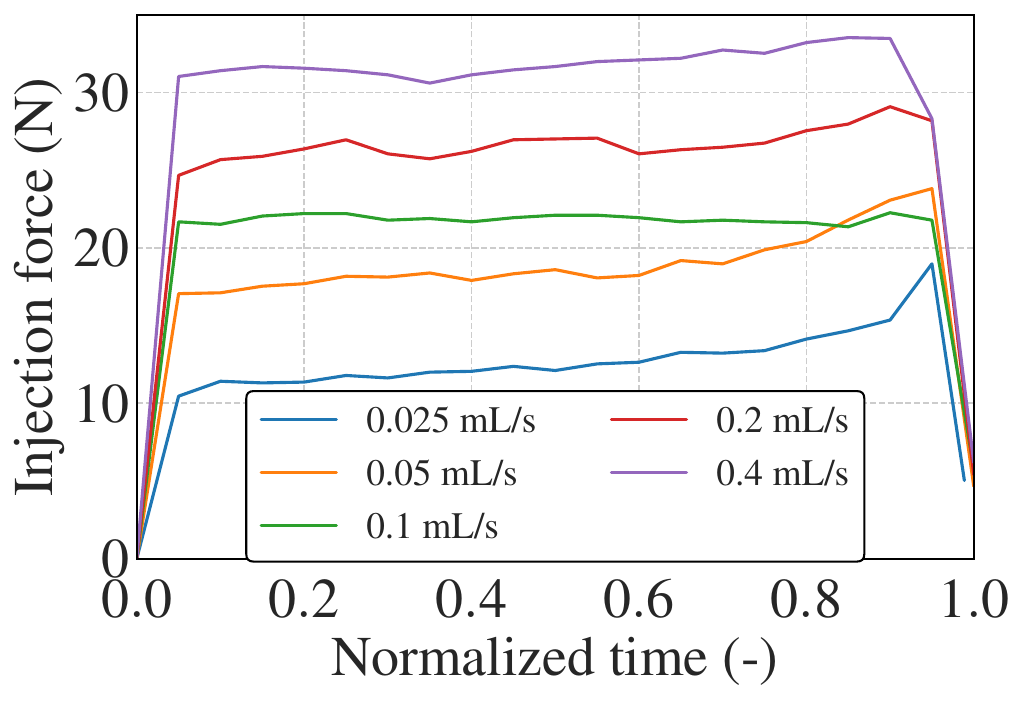}
		\caption{}
		\label{subfig:injforce}
	\end{subfigure}
	\begin{subfigure}{0.48\linewidth}
		\centering
		\includegraphics[width=\linewidth]{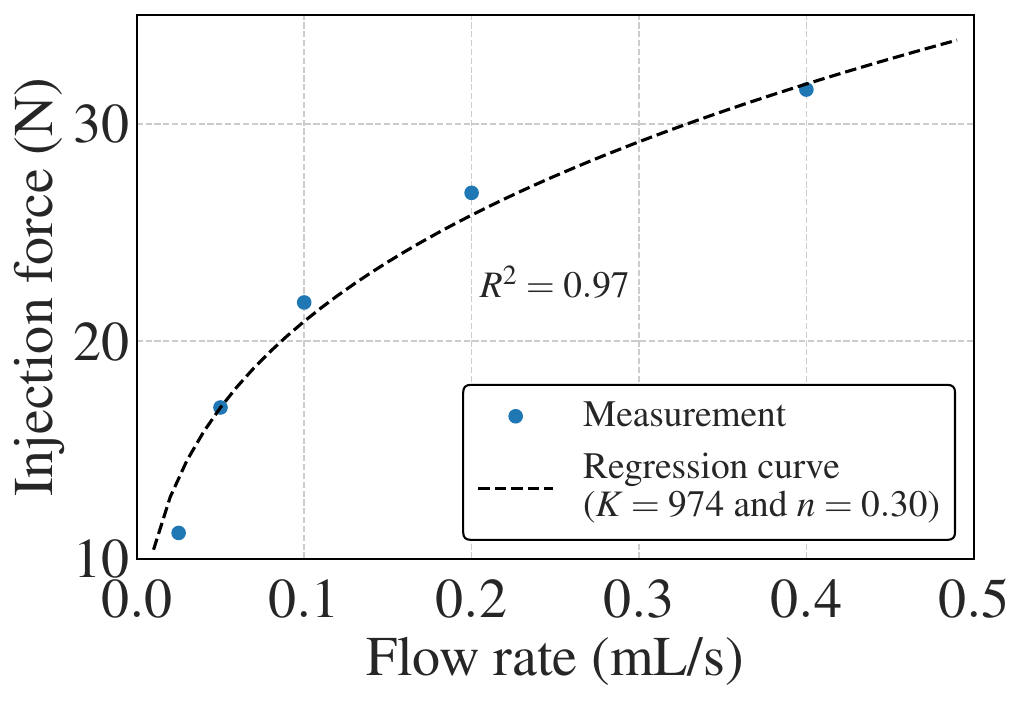}
		\caption{}
		\label{subfig:injforce_regression}
	\end{subfigure}
    \caption{(a) Injection force for various flow rates against time normalized by respective injection times (b) The average force values for each flow rate over a 5-second time period closest to the 245 seconds mark from the time of mixing were used for regression analysis using Equations \ref{eq:pres_ana2} and \ref{eq:force} to obtain power law parameters $K$ and $n$.}
    \label{fig:injforce}
\end{figure*}

The results of the injection experiments using the custom-made injector setup at various flow rates are presented in Figure \ref{fig:injforce}, wherein Figure \ref{subfig:injforce} shows the measured force against time normalized by the total duration of injection. As expected, the results showed that the higher the flow rate, the more injection force is required. For the two slowest flow rates, the force gradually increased. This was probably because the duration of injection was long enough and the shear rates were small enough for the cement to cure while being injected. To obtain reliable results, the average force values for each flow rate over a 5-second time period closest to the 245-second mark from the time of mixing were used for the regression analysis, as shown in Figure \ref{subfig:injforce_regression}. This was done to avoid any influence of curing time on the outcome of curve-fitting. The regression was carried out using Equations \ref{eq:pres_ana2} and \ref{eq:force} obtained from analytical calculations, with power law parameters $K$ and $n$ as fitting parameters. The regression analysis gave a good fit with $R^2=0.97$ yielding values $K = 974$ Pa s$^n$ and $n=0.30$. The value of $n$ is similar to those observed in test Rh7a for shear rate less than 1 s$^{-1}$ and in test Rh8 during the initial 30 seconds. We confirmed no difference between the fitting of power law and Herschley-Bulkey models in this case (refer Figure S4 in supporting information).

Using Equation \ref{eq:sr_r}, we could calculate the maximum shear rates, which were 10 s$^{-1}$ and 165 s$^{-1}$ for the lowest and the highest flow rates respectively. For shear rates in this range, we observed decreasing viscosity with time in test Rh8 (Figure \ref{subfig:visc_time}). However, if this was truly the case, it would cause a decrease in the injection force with time, which was not observed in our results. This is further evidence that the reducing viscosity seen in Figure \ref{subfig:visc_time} is due to the ridges formed in the sample as a result of the rotational deformation. In fact, surface degradation due to wall slip in injection flow conditions has been observed in previous studies \cite{tang_time-dependent_2008, tang_unsteady_2008}. However, there was no such evidence of wall slip in our experiments, as no difference in the surface consistencies was observed in the bone cement samples injected at different flow rates, despite the much higher shear rates compared to that of 0.082 s$^{-1}$ observed in the rotational rheometry. This is probably because the injection flow setting allows for much higher friction between the flowing bone cement and the walls as a result of the high injection pressures. The same was not possible in rotational rheometry with parallel plates, as even very small normal force caused the bone cement to squeeze out from between the parallel plates during our trials with normal force control. 

\subsection{Comparison of the methods}

\begin{table}
	\centering
	\begin{tabular}{l c c c}
		\toprule
		Test & Angular frequency * Strain amplitude/ & $K$ (Pa s$^n$) & $n$ \\ 
        &  Shear rate / Flow rate & & \\ \midrule
        Rh6, 0.01\% strain & 6.28 $\times$ 10$^{-5}$ to 0.0628 rad/s & 331 & 0.33 \\
        Rh6, 0.03\% strain & 1.88 $\times$ 10$^{-4}$ to 0.188 rad/s & 589 & 0.28 \\
		Rh6, 0.2\% strain & 0.00126 to 1.26 rad/s & 1060 & 0.27 \\
        %Rh6, 200\% strain & 1.26 to 1260 rad/s &  &  \\
		Rh7 & 0.001 to 1 s$^{-1}$ & 700-1100 & 0.20-0.30 \\
		Rh8 & 0.001 to 1 s$^{-1}$ & 1083 & 0.33 \\
		Inj1--Inj5 & 0.025 to 0.4 mL/s & 974 & 0.30  \\ 
		\bottomrule
	\end{tabular}
	\caption{Summary of power law parameters as obtained from various tests}
	\label{tab:summary} 
\end{table}

The values of the parameters of the power law equation $K$ and $n$ are useful for analyzing and simulating the injection flow behavior of bone cement. Usually, these values are extracted from experimentally measured flow curves using rheometer tests like test Rh7a, or the oscillatory frequency sweep test Rh6 if the Cox-Merz rule is known to be valid. \iffalse Our results from oscillatory frequency sweep (tests Rh6), rotational shear rate tests (tests Rh7, 7a, and 8), and injection experiments (tests Inj1 to Inj5) highlight the advantages and drawbacks of each of these testing methods. \fi They can also be extracted from a capillary rheometer or an injector setup like in tests Inj1--5. The power law parameters extracted from these tests are summarized in Table \ref{tab:summary}. 

Analytical calculations show that the shear rate range during an actual injection could go as high as 200 s$^{-1}$. As previously discussed, the rotational rheometer is not suitable at such high shear rates because of the wall sip and ridge formation caused by rotational deformation. This could be avoided by reducing the gap size between the rheometer plates, but that cannot be done in this case since the gap size must also be sufficiently large to ensure no particle effects. For the same reason, we could not use our cone-plate setup on the rheometer as it could only be used with a fixed gap of 49 microns, which is not large enough for our application. Hence, only measurements for shear rates below 1 s$^{-1}$ can be reliably used, as the power law parameters thus obtained are close to those obtained from the injection experiment. It must be noted, however, that even for low shear rates, the ridge formation could occur if the test duration was sufficiently long. It is important to exercise these cautions when performing rotational rheometer tests on PMMA bone cements. 

The flow curves can also be obtained from the oscillatory frequency sweep, but as seen from Figure \ref{fig:cox-merz}, it is important to ensure that the strain amplitude used for the frequency sweep lies within the non-linear viscoelastic region. Only then is the Cox-Merz rule valid and a feasible comparison to steady shear flow possible. This is a crucial finding, since this is in contrast to the conventional norm of performing the frequency sweep with a strain amplitude within the linear viscoelastic region. Note that the viscosity drop seen as a result of wall slip in the steady shear viscosity from rotational measurements is not seen in the oscillatory frequency sweep results. This, along with previously detailed visual and microscopical inspection, confirmed that the problem of wall slip or ridge formation is absent in oscillatory measurements. This demonstrates the advantage of oscillatory tests over rotational tests, especially as the shear rates go higher. 

The setup of a capillary rheometer or the custom-made injector setup used in this study is closest to an actual vertebroplasty setting. This kind of setup circumvents the drawbacks of the rotational or oscillatory rheometry, and yields mechanical parameters like effective or apparent viscosity and injection force that is most representative of the real-world conditions. It is important to note that the flow model used and the analytical calculations play an important role in the accuracy of the extracted rheological parameters. The Herschel-Buckley model is generally used for materials with yield stress. However, we did not observe significant difference in the fitting compared to the power law model (refer Figures S2, S3, and S4 in supporting information). Therefore, the power law model was used in this study for simplicity. As a drawback of this study, for an experimental setup involving injection through a circular tube or capillary, there need to be correction factors applied for flow entry, exit, change in tube diameter, etc. However, they have been ignored here since they are beyond the scope of this work. Hence, the rheological parameters extracted from our injector setup should be regarded as indicative and dependent on the specific setup and experimental conditions.

\section{Conclusion}

In this study, we conducted various tests using rotational rheometry, oscillatory rheometry, and a custom-made injector setup to understand the rheological behavior of a PMMA bone cement, here Vertecem V+, in the context of vertebroplasty. The bone cement exhibited low yield stress and fluid-like behavior under deformation, despite predominantly elastic contribution at rest, making it suitable for injection. Shear-thinning characteristics were observed, with the power law model parameters ranging between $K = 900 - 1100$ Pa s$^n$ and $n = 0.27 - 0.33$ across measurements. The two-phase curing process allowed the bone cement to be worked and injected during the initial slow curing phase, followed by the rapid curing phase. The curing rate of the bone cement was diminished upon deformation, while it rapidly accelerated upon exposure to the human body temperature. This needs to be considered when injecting bone cement in steps of smaller injections during vertebroplasty. Rotational rheometry was found to be susceptible to measurement artifacts like underestimation of viscosity, especially above 0.082 $s^{-1}$ shear rate. The cause was confirmed to be slippage between the bone cement sample and rheometer plates, and formation of circular ridges within the sample material. These measurement artifacts were not observed in oscillatory rheometry and injection tests. Finally, the Cox-Merz rule was found to be valid only for strain amplitudes in the non-linear viscoelastic region. This conditional validity of the Cox-Merz rule is critical when using oscillatory measurements to obtain the shear-thinning characteristics. Our findings underscore the impact of conditions and measurement methods on measured flow behavior. Understanding these nuances is crucial when interpreting bone cement rheology data for medical applications or their simulations.

\section*{Funding}
Funded by the Deutsche Forschungsgemeinschaft (DFG, German Research Foundation) – Project Number 327154368 – SFB 1313. We also acknowledge the support by the Stuttgart Center for Simulation Science (SimTech). J.K.W. acknowledges the European Union’s Horizon 2020 (H2020-MSCA-IF-2019) research and innovation programme under the Marie Sklodowska-Curie grant agreement 893099 — ImmunoBioInks.

\section*{CRediT statement}
\textbf{Zubin Trivedi}: Conceptualization, Methodology, Formal analysis, Investigation, Data curation, Validation, Writing - Original Draft, Writing – review \& editing, Visualization \textbf{Jacek K. Wychowaniec}: Conceptualization, Data curation, Formal analysis, Investigation, Methodology, Validation, Visualization, Writing – original draft, Writing – review \& editing \textbf{Dominic Gehweiler}: Methodology, Software, Resources, Project administration \textbf{Christoph M.~Sprecher}: Methodology, Investigation \textbf{Andreas Boger}: Formal analysis, Writing - Review \& Editing, Supervision \textbf{Boyko Gueorguiev}: Project administration, Supervision, Funding acquisition \textbf{Matteo D'Este}: Resources, Writing - Review \& Editing, Supervision, Project administration \textbf{Tim Ricken}: Supervision, Project administration, Funding acquisition \textbf{Oliver R\"ohrle}: Supervision, Project administration, Funding acquisition

% \begin{acknowledgment}

% Funded by the Deutsche Forschungsgemeinschaft (DFG, German Research Foundation) – Project Number 327154368 – SFB 1313. We also acknowledge the support by the Stuttgart Center for Simulation Science (SimTech). J.K.W. acknowledges the European Union’s Horizon 2020 (H2020-MSCA-IF-2019) research and innovation programme under the Marie Sklodowska-Curie grant agreement 893099 — ImmunoBioInks.
% \end{acknowledgment}

\begin{suppinfo}

% %This will usually read something like: ``Experimental procedures and
% %characterization data for all new compounds. The class will
% %automatically add a sentence pointing to the information on-line:
Additional figures containing the following supporting information are provided. 
\begin{itemize}
    \item Results of benchmark test at 3 mNm maximum torque, 1 Hz frequency, 23 °C on bone cement prepared by our mixing process
    \item Comparison of fitting using the Herschel-Bulkley and the power law rheological models on the rotational shear rate sweep test from 10$^{-4}$ to 10$^{-2}$ s$^{-1}$
    \item Comparison of fitting using the Herschel-Bulkley and the power law rheological models on the rotational shear rate sweep test from 10$^{-2}$ to 1 s$^{-1}$
    \item Comparison of fitting using the Herschel-Bulkley and the power law rheological models on the injection force measurement tests at various flow rates
\end{itemize}
Measurement data and test settings can be accessed from our data repository through the link \url{https://doi.org/10.18419/darus-4004}.

\end{suppinfo}

%\bibliographystyle{biochem}
\bibliography{references}

\end{document}